\documentclass[12pt]{article}
\usepackage{amsfonts, amsmath, amssymb, cite, graphicx, }

\usepackage{color}
\usepackage[all, knot]{xy}
\usepackage{tikz}
\usepackage{subfigure}
\usepackage{braket}

\usepackage[utf8]{inputenc}
\usepackage{ulem}

\usepackage{bm}
\usepackage{epstopdf}
\usepackage[footnotesize]{caption}
\usepackage{amsthm}
\usepackage{amsthm,amsmath,amssymb}
\usepackage{enumitem}
\usepackage{mathrsfs}
\usepackage{makecell}
\usepackage{diagbox}

\usepackage{pifont}

\usepackage[margin=3cm]{geometry}

\begin{document}

\begin{titlepage}
\vspace{0.5cm}
\begin{center}
{\Large \bf Thermal bath induced suppression of Hawking radiation}

\lineskip .75em
\vskip 2.5cm
{\large Hong Wang$^{a}$ and Jin Wang$^{b,}$\footnote{Corresponding author, jin.wang.1@stonybrook.edu} }
\vskip 2.5em
 {\normalsize\it $^{a}$State Key Laboratory of Electroanalytical Chemistry, Changchun Institute of Applied Chemistry, Chinese Academy of Sciences, Changchun 130022, China\\
 $^{b}$Department of Chemistry and Department of Physics and Astronomy, State University of New York at Stony Brook, NY 11794, USA}
\vskip 3.0em
\end{center}
\begin{abstract}
The cosmic microwave background radiation or  particles surrounding  a black hole can be modeled  as a thermal bath coupled to the black hole. In this study, we investigate the effects of such a thermal bath on black hole radiation. The model involves Hawking radiation particles that are bilinearly coupled to the thermal bath. In this framework, the Hawking radiation particles are treated as the system, and the bath serves as the environment. We analytically derive the effective Keldysh action and the dynamical equation of the system, which are essential for studying the effective dynamics of Hawking radiation. For the specific case of a bath weakly coupled to the radiation particles from a Schwarzschild black hole, we numerically solve the effective dynamical equation. The results indicate that the bath  suppresses Hawking radiation. Consequently, an observer located far away  from the black hole will detect radiation that is  weaker than Hawking’s original prediction.
\end{abstract}
\end{titlepage}

\baselineskip=0.7cm

\tableofcontents
\newpage
\section{Introduction}
\label{sec:1}

Black hole radiation has been extensively studied for many years~\cite{SW1,SW2,JSW,DP,PF,KT,STK,AM,EC,RS,SG,CJF,AND,TS,TR,ZJ,HYJ,CKD}. Although it has not yet been experimentally observed, researchers  generally believe that Hawking radiation should be present,  as it can be derived through several independent approaches.  Hawking's original method is based on  quantum field theory in curved spacetime, in which he showed that  black hole radiation possesses the characteristics of a blackbody spectrum~\cite{SW2}. In Ref.~\cite{MF},  Parikh and Wilczek used the quantum tunneling method to show that Hawking radiation may exhibit small deviations from a perfect blackbody spectrum. The quantum tunneling method helps to indicate that  black hole radiation obeys the principles of quantum mechanics, and it has attracted considerable attention in subsequent studies~\cite{EVD,ETD,QXX,RB,MML,JZ,JZ2,JZ3,SQ,PM,BQM,XW,MLS}. Recent developments involving the black hole island formula have more clearly demonstrated that black hole radiation does not violate the unitarity of quantum mechanics~\cite{ATJ13,ARJ,G13,TEA13,KTA13,GSD13}.  Furthermore, some studies suggest that black hole radiation is necessary to prevent the breakdown of general covariance at the quantum level~\cite{SHF,SF}.  Since black hole radiation is a type of quantum effect. It may play an important role in our understanding of quantum gravity.

In the actual universe, black holes  can accrete  interstellar particles, resulting in a distribution of  various particles around them. This scenario is somewhat analogous to Earth's atmosphere. We know that the atmosphere can be regarded as a thermal bath (with temperature varying at different altitudes ) that is coupled to the Earth. If a particle beam is launched from the Earth's surface into outer space, the atmosphere may influence  its dynamics. Similarly, the particles surrounding a black hole can be approximately treated as a thermal bath. It is natural to expect that when  particles emitted  from the black hole horizon pass through this bath, the bath will interact with the Hawking radiation particles (HRPs). Therefore, it is interesting to study how the bath surrounding a black hole affects the properties of HRPs.  Additionally, some researchers have pointed out that the cosmic microwave background (CMB) can also be viewed as a thermal bath that  couples to the black hole~\cite{IAA}.

Several studies have investigated  the impact of a thermal bath on the  black hole. In Ref.~\cite{PJ}, Cust$\mathrm{\acute{o}}$dio and Horvath showed that when a black hole is coupled to  a bath, it may not evaporate completely via Hawking radiation, potentially leaving a Planck scale remnant as the final state of a primordial black hole.  In Ref.~\cite{HS}, Saida showed  that the evaporation timescale of a black hole in a bath can differ from that in  empty space. In Ref.~\cite{IAA}, Agullo, Brady, Delhom and Kranas demonstrated that the presence of a thermal bath can suppress entanglement between Hawking radiation and the black hole. In Ref.~\cite{AJD12}, Chatterjee, Kalita and Maity showed that when a black hole is coupled to a finite temperature thermal bath, the black hole radiation includes both spontaneous and stimulated emission. The spontaneous emission corresponds to conventional Hawking radiation, while the stimulated emission causes the radiation spectrum to deviate from a blackbody spectrum and enhances the evaporation rate of the black hole.  Moreover, to study the black hole information paradox, several researchers  have  considered  models in which a black hole is coupled to a bath and have shown that the evolution of the von Neumann entropy of the Hawking radiation follows the Page curve~\cite{ARJ,HZJ}.  Other significant studies have also been conducted; see Refs.~\cite{JM,BS,HY,DC,JDA12} and references therein. We will not enumerate all of them here.

In the models considered in Refs.~\cite{IAA,AJD12,ARJ,HZJ}, the HRPs and the  bath are described by the same field. However, a black hole can radiate various  types of particles. Therefore, the HRPs and the paticles in the bath do not necessarily belong to the same field quanta. For example, when a black hole is coupled to the CMB bath, the bath particles are, strictly speaking, quanta of a vector field, whereas the particles radiated from the black hole may be scalar field quanta or quanta of other fields. It is thus important  to study scenarios in which the bath and the HRPs are described by different fields.

In this work, we try to study the effective quantum dynamics of  the HRPs when the black hole  coupled to a bath, with the HRPs and the bath described by different fields. To our knowledge, this remains an unresolved issue.  For simplicity, the interaction between the HRP and the bath is assumed to be  bilinear.   In the theory of open quantum systems, it is  common  to identify the degrees of freedom of interest as the system and treat the remaining  degrees  as the environment~\cite{HF}. Since we are interested in the dynamics of the HRPs, they are regarded as an open quantum system, with the bath serving as the environment.  We analytically derive the effective Keldysh action and the dynamical equation of the system, which are essential for investigating the effective dynamics of the HRP in the presence of an environment.

The resulting effective dynamical equation obtained is an integro-differential equation.  Compared to the traditional Klein-Gordon equation for scalar fields in curved spacetime, the effective dynamical equation describing the HRP in our model includes an additional  term that captures the influence of the bath on the HRP's dynamics. We show that the magnitude of this term depends on the position in curved spacetime, whereas it remains constant in flat spacetime. In the limit of weak coupling and small black hole mass, the four dimensional effective dynamical equation can be reduced to a one dimensional inhomogeneous differential equation. In this simplified case, we numerically solve the dynamical equation. The results indicate that the bath suppresses Hawking radiation. Thus, if a black hole is surrounded by a bath, an observer far from the black hole will detect  radiation weaker than  that predicted by Hawking’s original result. Throughout this study, we work in units where $G=\hbar=k_{B}=c=1$.

\section{The model }
\label{sec:2}

For simplicity, by neglecting spin and other internal degrees of freedom, both the HRP and the bath can be described by scalar fields~\cite{SW2,LD,HJ,HJ2}, which  is a typical effective approach in field theory. The scalar fields $\phi$ and $\chi$  are used to represent the HRP and the bath, respectively.   For simplicity, the interaction between the two scalar fields is assumed to be  bilinear. In this case, the action of the total system (system+environment) under curved spacetime background  can be written as~\cite{SSS}
\begin{eqnarray}\begin{split}
\label{eq:2.1}
S_{tot}=&\frac{1}{2}\int d^{4}x \sqrt{-g} \big(g^{\mu\nu}\partial_{\mu}\phi\partial_{\nu}\phi-m_{\phi}^{2}\phi^{2}\big)+\frac{1}{2}\int d^{4}x \sqrt{-g} \big(g^{\mu\nu}\partial_{\mu}\chi\partial_{\nu}\chi-m_{\chi}^{2}\chi^{2}\big)\\&+\alpha\int d^{4}x \sqrt{-g}\phi\chi.
\end{split}
\end{eqnarray}
Here,  $S_{tot}$ denotes the action of the total system. $m_{\phi}$ and $m_{\chi}$ are the masses of the scalar fields $\phi$ and $\chi$, respectively. $g_{\mu\nu}$ is the spacetime metric, and $\alpha$ is the coupling constant between the two fields.

It is clear  that the first and second terms on the right hand side of Eq. \eqref{eq:2.1} correspond to  the actions of the scalar field $\phi$ and $\chi$ under curved spacetime background,  denoted as $S_{\phi}$ and $S_{\chi}$, respectively. The third term represents the interaction, denoted as $S_{int}$. Thus, the total action can  be expressed as $S_{tot}=S_{\phi}+S_{\chi}+S_{int}$.  The model defined by Eq. \eqref{eq:2.1} can be viewed as a generalization of the Caldeira-Leggett model to curved spacetime~\cite{SSS}.  In other words, the Caldeira-Leggett model can be regarded as the non-relativistic limit of the model defined by Eq. \eqref{eq:2.1}.  The Caldeira-Leggett model  describes the influence of a bath on the quantum dynamics of a non-relativistic particle~\cite{AA}.  The interaction between an HRP and the bath surrounding a black hole is the curved spacetime analogue of a non-relativistic particle interacting with a bath.  Therefore, it is reasonable to use the model in Eq. \eqref{eq:2.1} to describe the effect of the thermal bath on the HRPs.

For simplicity, we consider the case where the spacetime admits a timelike Killing vector field. That is, the metric $g_{\mu\nu}$  is independent of the time coordinate, and the spacetime is stationary. In this case,  a conserved Hamiltonian  can be defined for the total system. From Eq. \eqref{eq:2.1},  the Lagrangian density and  Hamiltonian density of the total system are derived as
\begin{eqnarray}\begin{split}
\label{eq:2.2}
\mathscr{L}_{tot}=\frac{1}{2}\sqrt{-g} \big(g^{\mu\nu}\partial_{\mu}\phi\partial_{\nu}\phi-m_{\phi}^{2}\phi^{2}\big)+\frac{1}{2} \sqrt{-g} \big(g^{\mu\nu}\partial_{\mu}\chi\partial_{\nu}\chi-m_{\chi}^{2}\chi^{2}\big)+\alpha \sqrt{-g}\phi\chi,
\end{split}
\end{eqnarray}
\begin{eqnarray}\begin{split}
\label{eq:2.3}
\mathscr{H}_{tot}=&\frac{1}{2} \sqrt{-g} \big(g^{00}\dot{\phi}^{2}-g^{ii}(\partial_{i}\phi)^{2}+m_{\phi}^{2}\phi^{2}\big)\\&+\frac{1}{2} \sqrt{-g} \big(g^{00}\dot{\chi}^{2}-g^{ii}(\partial_{i}\chi)^{2}+m_{\chi}^{2}\chi^{2}\big)-\alpha\sqrt{-g}\phi\chi,
\end{split}
\end{eqnarray}
where, $\dot{\phi}\equiv d\phi/dt$, $\dot{\chi}\equiv d\chi/dt$,  and $t$ represents the time coordinate. The Latin indexes $i$, $j$... $= 1, 2, 3$ are reserved for spatial  coordinates, while the Greek indexes $\mu$, $\nu$... $= 0, 1, 2, 3$ encompass the spacetime coordinates.

The relationship between $\mathscr{L}_{tot}$ and $S_{tot}$ is $S_{tot}=\int d^{4}x\mathscr{L}_{tot}$, while  the Hamiltonian density is related to the Lagrangian density via the Legendre transformation: $\mathscr{H}_{tot}=\pi_{\phi}\dot{\phi}+\pi_{\chi}\dot{\chi}-\mathscr{L}_{tot}$,  where, $\pi_{\phi}$ and $\pi_{\chi}$ are the conjugate momenta of $\phi$ and $\chi$, respectively. The first and second  terms on the right hand side of Eq. \eqref{eq:2.2} correspond to the Lagrangian densities of the field $\phi$  and $\chi$,  denoted as $\mathscr{L}_{\phi}$  and $\mathscr{L}_{\chi}$. The third term is the interaction Lagrangian density  $\mathscr{L}_{int}$. Similarly, on the right hand side of Eq. \eqref{eq:2.3}, the first and second terms   are the Hamiltonian densities of  $\phi$ and $\chi$, denoted as $\mathscr{H}_{\phi}$ and $\mathscr{H}_{\chi}$, and the third term is the interaction Hamiltonian density  $\mathscr{H}_{int}$. Thus, the total Lagrangian and  Hamiltonian densities  can be written as $\mathscr{L}_{tot}=\mathscr{L}_{\phi}+\mathscr{L}_{\chi}+\mathscr{L}_{int}$ and $\mathscr{H}_{tot}=\mathscr{H}_{\phi}+\mathscr{H}_{\chi}+\mathscr{H}_{int}$, respectively. The corresponding Hamiltonians are defined as: $H_{tot}\equiv \int d^{3}x \mathscr{H}_{tot}$, $H_{\phi}\equiv \int d^{3}x \mathscr{H}_{\phi}$, $H_{\chi}\equiv \int d^{3}x \mathscr{H}_{\chi}$, and $H_{int}\equiv \int d^{3}x \mathscr{H}_{int}$. From these, one can easily show that
\begin{equation}
\label{eq:2.4}
H_{tot}=H_{\phi}+H_{\chi}+H_{int},
\end{equation}
\begin{equation}
\label{eq:2.5}
H_{\phi}=\frac{1}{2}\int d^{3}x \big\{\frac{\pi_{\phi}^{2}}{\sqrt{-g}g^{00}}-\sqrt{-g}g^{ii}(\partial_{i}\phi)^{2}+\sqrt{-g}m_{\phi}^{2}\phi^{2}\big\} ,
\end{equation}
\begin{equation}
\label{eq:2.6}
H_{\chi}= \frac{1}{2}\int d^{3}x\big \{\frac{\pi_{\chi}^{2}}{\sqrt{-g}g^{00}}-\sqrt{-g}g^{ii}(\partial_{i}\chi)^{2}+\sqrt{-g}m_{\chi}^{2}\chi^{2}\big\},
\end{equation}
\begin{equation}
\label{eq:2.7}
H_{int}= -\alpha \int d^{3}x\sqrt{-g}\phi\chi.
\end{equation}

Both the HRPs and the bath possess nonzero energy-momentum tensors. According to general relativity, these tensors  will affect the spacetime structure. However, the influence of  HRPs on  spacetime  is typically negligible. For the bath, if  its effect on the spacetime structure is taken into account,   Hawking radiation could be modified by a greybody factor~\cite{QWR,SI}. For simplicity,  we restrict our analysis to the regime in which the impact of the bath on the spacetime metric is sufficiently small to be neglected. Thus, the influence of the bath on the HRPs arises solely from the interaction term in  Eq. \eqref{eq:2.1}.

\section{Effective Keldysh action and dynamical equation}
\label{sec:3}

Upon quantizing  the total system, its quantum dynamics can be obtained. For notational simplicity, we omit the operator hats. Readers can readily distinguish between $c$-numbers and $q$-numbers from the context. For an open quantum system, the complete dynamical information is encoded in the reduced density matrix, whose evolution is governed by the Liouville–von Neumann equation~\cite{HF,HJC}
\begin{equation}
\label{eq:3.1}
\frac{d\rho}{dt}=-i\mathrm{Tr}_{\chi}[H_{tot},\rho_{tot}].
\end{equation}
Here, $\rho$ is the reduced density matrix of the system,  $\rho_{tot}$ is the total density matrix of the  system and bath, and $\mathrm{Tr}_{\chi}$  denotes the partial trace over the bath.

\begin{figure}[tbp]
\centering
\includegraphics[width=10cm]{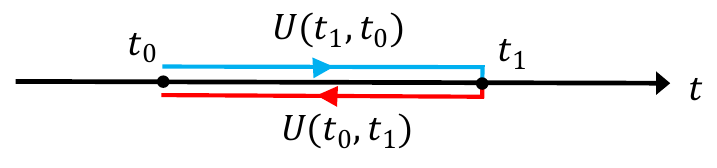}
\caption{\label{fig:a}  The closed time path contour. The blue arrowed line represents the forward time path, whereas the red arrowed line represents the backward time path. }
\end{figure}

Equation \eqref{eq:3.1} can also be written as
\begin{equation}
\label{eq:3.2}
\rho(t)=\mathrm{Tr}_{\chi}\big(U(t,t_{0})\rho_{tot}(t_{0})U(t_{0},t)\big),
\end{equation}
where $t_{0}$ represents the initial time.  $U(t, t_{0})$ is the time  evolution operator from the initial time $t_{0}$ to  time $t$, and it satisfies   the property $U(t_{0},t)=U^{\dag}(t,t_{0})$. The operator $U(t, t_{0})$ corresponds to the forward time path $t_{0}\rightarrow t$   , while $U(t_{0},t)$ corresponds to the backward time path $t\rightarrow t_{0}$, as shown in Fig.~\ref{fig:a}.   The relation between  $U(t,t_{0})$  and the total Hamiltonian operator is
\begin{equation}
\label{eq:3.3}
U(t,t_{0})=\mathbf{T}\:\mathrm{exp}\{-i\int_{t_{0}}^{t}H_{tot}dt\}=e^{-iH_{tot}\delta t}\cdot\cdot\cdot e^{-iH_{tot}\delta t},
\end{equation}
where $\mathbf{T}$ is the chronological time ordering operator and  $\delta t$ denotes an infinitesimal time interval. From the first  to the second step, the time variable is discretized ($t\rightarrow t_{0},t_{1}, t_{2},\cdot\cdot\cdot, t_{N-1}, t_{N}=t$, with $\delta t=t_{n}-t_{n-1} $).   At the initial time, it is often assumed that the system and bath are unentangled~\cite{HF}. Under this assumption, the initial density matrix of the total system is given by $\rho_{tot}(t_{0})=\rho(t_{0})\otimes \rho_{\chi}(t_{0})$, where $\rho_{\chi}(t_{0})$ is the initial density matrix of the bath. Substituting this assumption and Eq. \eqref{eq:3.3} into Eq. \eqref{eq:3.2} and taking $t_{0}=-\infty$, the density matrix at time $t_{N}$  can be written as
\begin{equation}
\label{eq:3.4}
\rho(t_{N})=\mathrm{Tr}_{\chi}\{e^{-iH_{tot}\delta t}\cdot\cdot\cdot e^{-iH_{tot}\delta t}\rho(-\infty)\otimes \rho_{\chi}(-\infty)e^{iH_{tot}\delta t}\cdot\cdot\cdot e^{iH_{tot}\delta t}\}.
\end{equation}
Equations \eqref{eq:3.2} and \eqref{eq:3.4} indicate that the evolution of the density matrix depends on both the forward and backward time paths, which is a  characteristic feature  of the non-equilibrium or non-unitary dynamics of an open quantum system~\cite{JR,AK}.

To express the density matrix $\rho$ in the path integral formalism, we insert  the identities $\int d\phi_{+n}|\phi_{+n}\rangle\langle\phi_{+n}|=\mathbf{1}$ and $\int d\phi_{-n}|\phi_{-n}\rangle\langle\phi_{-n}|=\mathbf{1}$ ( $|\phi\rangle$ denotes an eigenstate of the field operator $\phi$) into Eq. \eqref{eq:3.4},  yielding
\begin{eqnarray}\begin{split}
\label{eq:3.5}
\rho(t_{N})=&\mathrm{Tr}_{\chi}\Big\{e^{-iH_{tot}\delta t}\int d\phi_{+(N-1)}|\phi_{+(N-1)}\rangle\langle\phi_{+(N-1)}|e^{-iH_{tot}\delta t}\int d\phi_{+(N-2)}|\phi_{+(N-2)}\rangle\cdot\cdot\cdot\\&e^{-iH_{tot}\delta t}\int d\phi_{+1}|\phi_{+1}\rangle\langle\phi_{+1}|\rho(-\infty)\otimes \rho_{\chi}(-\infty)\int d\phi_{-1}|\phi_{-1}\rangle\langle\phi_{-1}|e^{iH_{tot}\delta t}\\&\int d\phi_{-2}|\phi_{-2}\rangle\langle\phi_{-2}|e^{iH_{tot}\delta t}\cdot\cdot\cdot\int d\phi_{-(N-1)}|\phi_{-(N-1)}\rangle\langle\phi_{-(N-1)}|e^{iH_{tot}\delta t}\Big\}.
\end{split}
\end{eqnarray}
Quantities labeled with subscripts  $+$ and $-$ correspond to the forward and backward time paths, respectively.

Let  $|\pi_{\phi}\rangle$ denote an eigenstate of the momentum operator $\pi_{\phi}$. Using the relation  $\langle\phi|\pi_{\phi}\rangle=e^{i\phi\pi_{\phi}}$, one finds
\begin{eqnarray}\begin{split}
\label{eq:3.6}
\langle\phi_{+n}|e^{-iH_{tot}\delta t}|\phi_{+(n-1)}\rangle=\int d\pi_{\phi_{+n}}\:&\mathrm{exp}\Big\{i\big(\pi_{\phi_{+n}}\dot{\phi}_{+n}-H_{\phi}(\pi_{\phi_{+n}},\phi_{+(n-1)})\big)\delta t\Big\}\\&\times\mathrm{exp}\Big\{-i\big(H_{\chi}+H_{int}(\phi_{+(n-1)})\big)\delta t\Big\},
\end{split}
\end{eqnarray}
and
\begin{eqnarray}\begin{split}
\label{eq:3.7}
\langle\phi_{-n}|e^{iH_{tot}\delta t}|\phi_{-(n+1)}\rangle=\int d\pi_{\phi_{-n}}\:&\mathrm{exp}\Big\{-i\big(\pi_{\phi_{-n}}\dot{\phi}_{-(n+1)}-H_{\phi}(\pi_{\phi_{-n}},\phi_{-(n+1)})\big)\delta t\Big\}\\&\times\mathrm{exp}\Big\{i\big(H_{\chi}+H_{int}(\phi_{-(n+1)})\big)\delta t\Big\}.
\end{split}
\end{eqnarray}
Here, $\pi_{\phi_{+n}}$ ( $\pi_{\phi_{-n}}$) denotes the conjugate momentum of $\phi_{+n}$ ($\phi_{-n}$). $H_{\phi}(\pi_{\phi_{+n}},\phi_{+(n-1)})$ and $H_{int}(\phi_{+(n-1)})$ are given by
\begin{equation}
\label{eq:3.8}
H_{\phi}(\pi_{\phi_{+n}},\phi_{+(n-1)})=\frac{1}{2}\int d^{3}x \Big\{\frac{\pi_{\phi_{+n}}^{2}}{\sqrt{-g}g^{00}}-\sqrt{-g}g^{ii}(\partial_{i}\phi_{+(n-1)})^{2}+\sqrt{-g}m_{\phi}^{2}\phi_{+(n-1)}^{2}\Big\}
\end{equation}
and
\begin{equation}
\label{eq:3.9}
H_{int}(\phi_{+(n-1)})= -\alpha \int d^{3}x\sqrt{-g}\phi_{+(n-1)}\chi.
\end{equation}
Upon replacing $\pi_{\phi}$ and $\phi$ with $\pi_{\phi_{+n}}$ and $\phi_{+(n-1)}$, respectively, Eqs. \eqref{eq:2.5} and \eqref{eq:2.7} are transformed into  Eqs. \eqref{eq:3.8} and \eqref{eq:3.9}. The definitions of $H_{\phi}(\pi_{\phi_{-n}},\phi_{-(n+1)})$ and $H_{int}(\phi_{-(n+1)})$ are analogous.

The  density matrix element at time $t_{N}$ is defined as $\rho(\phi_{+N},\phi_{-N})\equiv\langle \phi_{+N}|\rho(t_{N})| \phi_{-N}\rangle$. Combining  Eqs. \eqref{eq:3.5}-\eqref{eq:3.7},  $\rho(\phi_{+N},\phi_{-N})$ can be written as
\begin{eqnarray}\begin{split}
\label{eq:3.10}
\rho(\phi_{+N},\phi_{-N})=&\int d\phi_{+(N-1)}\cdot\cdot\cdot d\phi_{+1}d\phi_{-(N-1)}\cdot\cdot\cdot d\phi_{-1}d\pi_{\phi_{+N}}\cdot\cdot\cdot d\pi_{\phi_{+2}}d\pi_{\phi_{-1}}\cdot\cdot\cdot d\pi_{\phi_{-(N-1)}}\\&\times \mathrm{exp}\Big\{i\delta t\sum_{n}\big(\pi_{\phi_{+n}}\dot{\phi}_{+n}-H_{\phi}(\pi_{\phi_{+n}},\phi_{+(n-1)})\big)\Big\}\langle\phi_{+1}|\rho(-\infty)|\phi_{-1}\rangle\\&\times \mathrm{exp}\Big\{-i\delta t\sum_{n}\big(\pi_{\phi_{-n}}\dot{\phi}_{-(n+1)}-H_{\phi}(\pi_{\phi_{-n}},\phi_{-(n+1)})\big)\Big\}\\&\times \mathrm{Tr}_{\chi}\Big\{\mathrm{exp}\big\{-i\delta t \sum_{n}(H_{\chi}+H_{int}(\phi_{+(n-1)}))\big\}\rho_{\chi}(-\infty)\\& \quad \quad\quad  \times\mathrm{exp}\big\{i\delta t \sum_{n}(H_{\chi}+H_{int}(\phi_{-(n+1)}))\big\}\Big\}.
\end{split}
\end{eqnarray}

Taking the continues limit $\delta t\rightarrow 0$ and rewriting  $\rho(\phi_{+N},\phi_{-N})$ as $\rho(\phi_{+},\phi_{-}; t)$, Eq. \eqref{eq:3.10} becomes
\begin{eqnarray}\begin{split}
\label{eq:3.11}
\rho(\phi_{+},\phi_{-};t)=&\int D \phi_{+}D \phi_{-}D \pi_{\phi_{+}}D \pi_{\phi_{-}}\:\mathrm{exp}\Big\{i\int dt (\pi_{\phi_{+}}\dot{\phi}_{+}-H_{\phi}(\pi_{\phi_{+}},\phi_{+})) \Big\}\\&\times\langle\phi_{+1}|\rho(-\infty)|\phi_{-1}\rangle\: \mathrm{exp}\Big\{-i\int dt (\pi_{\phi_{-}}\dot{\phi}_{-}-H_{\phi}(\pi_{\phi_{-}},\phi_{-}))\Big\}\mathbf{I}(\phi_{+},\phi_{-}).
\end{split}
\end{eqnarray}
Here,
 \begin{equation}
\label{eq:3.12}
\int D \phi_{+}\equiv \lim_{N\rightarrow\infty}\int d\phi_{+1}d\phi_{+2}\cdot\cdot\cdot d\phi_{+(N-1)},
\end{equation}
and
\begin{equation}
\label{eq:3.13}
\mathbf{I}(\phi_{+},\phi_{-})\equiv \mathrm{Tr}_{\chi}\Big\{\mathrm{exp}\big\{-i\int dt (H_{\chi}+H_{int}(\phi_{+}))\big\}\rho_{\chi}(-\infty)  \:\mathrm{exp}\big\{i\int dt (H_{\chi}+H_{int}(\phi_{-}))\big\}\Big\}.
\end{equation}
The definitions of $\int D \phi_{-}$, $\int D \pi_{\phi_{+}}$ and $\int D \pi_{\phi_{-}}$ are analogous  to that of $\int D \phi_{+}$.  The quantity  $\mathbf{I}(\phi_{+},\phi_{-})$ is the Feynman-Vernon  influence functional~\cite{HJ,ND,BJY,RF}, which describes the effect of the bath on the HRP.

Equation \eqref{eq:3.11} is the path integral representation of the density matrix element $\rho(\phi_{+},\phi_{-};t)$. After performing  the integrations over the conjugate momenta~\cite{XYZ}, Eq. \eqref{eq:3.11} reduces to
\begin{eqnarray}\begin{split}
\label{eq:3.14}
\rho(\phi_{+},\phi_{-};t)=&\int D \phi_{+}D \phi_{-}\:\mathrm{exp}\big\{iS_{\phi}^{K}\big\}\langle\phi_{+1}|\rho(-\infty)|\phi_{-1}\rangle \:\mathbf{I}(\phi_{+},\phi_{-}),
\end{split}
\end{eqnarray}
where
 \begin{equation}
\label{eq:3.15}
S_{\phi}^{K}\equiv\int d^{4}x (\mathscr{L}_{\phi}(\phi_{+})-\mathscr{L}_{\phi}(\phi_{-}))
\end{equation}
is the Keldysh action of the free scalar field $\phi$ in curved spacetime. In Eq. \eqref{eq:3.15},
\begin{equation}
\label{eq:3.16}
\mathscr{L}_{\phi}(\phi_{+})=\frac{1}{2}\sqrt{-g} \big(g^{\mu\nu}\partial_{\mu}\phi_{+}\partial_{\nu}\phi_{+}-m_{\phi}^{2}\phi_{+}^{2}\big).
\end{equation}
Replacing $\phi_{+}$ with $\phi_{-}$ gives  $\mathscr{L}_{\phi}(\phi_{-})$.

It is well known that the temperature of the Earth's atmosphere varies with altitude. By analogy, one may expect that the temperature of the bath surrounding a black hole  also varies with positions. However, as a toy model, we assume that the bath has a uniform temperature independent of spacetime coordinates. This assumption implies that the bath is in a stationary state, which is widely adopted in studies of black holes interacting with external bath.   Specifically, we take the bath state to be~\cite{HF}
\begin{equation}
\label{eq:3.17}
\rho_{\chi}\propto \mathrm{exp}\{-\frac{H_{\chi}}{T_{\chi}}\},
\end{equation}
where $T_{\chi}$ denotes the temperature of the bath.  Since our goal is to analyze the qualitative behavior of the HRP, we omit the normalization constant.

Substituting Eq. \eqref{eq:3.17} into Eq. \eqref{eq:3.13}, one can verify that the influence functional can be expressed as
\begin{equation}
\label{eq:3.18}
\mathbf{I}(\phi_{+},\phi_{-})=\mathrm{exp}\Big\{-\frac{\alpha^{2}}{2}\int d^{4}x d^{4}y \sqrt{-g(x)}\sqrt{-g(y)}\vec{\phi}(x)\mathbf{G}(x,y)\vec{\phi}^{\mathrm{T}}(y) \Big\},
\end{equation}
where $\vec{\phi}(x)=(\phi_{+}(x), \phi_{-}(x))$ and $\vec{\phi}^{\mathrm{T}}$ is its transpose. The matrix Green function $\mathbf{G}(x,y)$ is defined as~\cite{JR,AK}
\begin{equation}
\label{eq:3.19}
\mathbf{G}(x,y)\equiv
\begin{pmatrix}
G_{++}, & G_{+-}\\
G_{-+}, & G_{--}
\end{pmatrix}
\equiv
\begin{pmatrix}
\mathrm{Tr}\big(\rho_{\chi}\mathbf{T}\big(\chi(x)\chi(y)\big)\big), & -\mathrm{Tr}\big(\rho_{\chi}\chi(y)\chi(x)\big)\\
-\mathrm{Tr}\big(\rho_{\chi}\chi(x)\chi(y)\big), & \mathrm{Tr}\big(\rho_{\chi}\mathbf{\tilde{T}}\big(\chi(y)\chi(x)\big)\big)
\end{pmatrix},
\end{equation}
where, $\mathbf{\tilde{T}}$ is the anti-chronological time ordering operator. The derivation of Eq. \eqref{eq:3.18} is provided in Appendix~\ref{sec:A}.  The influence functional in Eq. \eqref{eq:3.18} is valid only under the following conditions: (\romannumeral1) the interaction between the HRP and the bath is bilinear, and (\romannumeral2) the bath is in the thermal equilibrium state given by Eq. \eqref{eq:3.17}.  See Appendix~\ref{sec:A} for details.   From the definitions in equation \eqref{eq:3.19}, we see that $-iG_{++}$ is the Feynman propagator, $iG_{--}$ is the Dyson propagator, while $iG_{+-}$ and $iG_{-+}$ are the Wightman functions~\cite{EB}.

Substituting Eq. \eqref{eq:3.18} into Eq. \eqref{eq:3.14}, the density matrix element $\rho(\phi_{+},\phi_{-};t)$ becomes
\begin{eqnarray}\begin{split}
\label{eq:3.20}
\rho(\phi_{+},\phi_{-};t)=&\int D \phi_{+}D \phi_{-}\:\mathrm{exp}\big\{iS_{eff}^{K}\big\}\langle\phi_{+1}|\rho(-\infty)|\phi_{-1}\rangle,
\end{split}
\end{eqnarray}
where
\begin{eqnarray}\begin{split}
\label{eq:3.21}
S_{eff}^{K}=&\int d^{4}x (\mathscr{L}_{\phi}(\phi_{+})-\mathscr{L}_{\phi}(\phi_{-}))+\frac{i\alpha^{2}}{2}\int d^{4}x d^{4}y \sqrt{-g(x)}\sqrt{-g(y)}\vec{\phi}(x)\mathbf{G}(x,y)\vec{\phi}^{\mathrm{T}}(y)
\end{split}
\end{eqnarray}
is the effective Keldysh action of the system.  The  dynamics of the HRP is governed by $S_{eff}^{K}$. Equation \eqref{eq:3.20} establishes the relationship between the initial and  final states of  the system.   On the right hand side of Eq. \eqref{eq:3.21}, the first term is the Keldysh action of the free scalar field $\phi$ in curved spacetime, while the second term represents the influence of the bath.

To study the semiclassical dynamics of the HRP, it is convenient to perform the Keldysh rotation of the action \eqref{eq:3.21}~\cite{JR,AK,LMS}:
\begin{eqnarray}
\label{eq:3.22}
 \begin{cases}
       \phi_{cl}\equiv\frac{1}{\sqrt{2}}(\phi_{+}+\phi_{-}), \\
       \phi_{q}\equiv\frac{1}{\sqrt{2}}(\phi_{+}-\phi_{-}),
  \end{cases}
\end{eqnarray}
where $\phi_{cl}$ and $\phi_{q}$ represent the classical and quantum components of the field $\phi$, respectively. Substituting Eq. \eqref{eq:3.22} into Eq. \eqref{eq:3.21}, the effective Keldysh action becomes
\begin{eqnarray}\begin{split}
\label{eq:3.23}
S_{eff}^{K}=&\int d^{4}x \sqrt{-g(x)}(g^{\mu\mu}\partial_{\mu}\phi_{q}\partial_{\mu}\phi_{cl}-m_{\phi}^{2}\phi_{q}\phi_{cl}) \\&+\frac{i\alpha^{2}}{4}\int d^{4}x d^{4}y \sqrt{-g(x)}\sqrt{-g(y)}(\phi_{cl}(x),\phi_{q}(x))\mathbf{G^{\circ}}(x,y)(\phi_{cl}(y),\phi_{q}(y))^{\mathrm{T}}.
\end{split}
\end{eqnarray}
Here,
\begin{equation}
\label{eq:3.24}
\mathbf{G^{\circ}}(x,y)\equiv
\begin{pmatrix}
G_{++}+G_{-+}+G_{+-}+G_{--}, & G_{++}+G_{-+}-G_{+-}-G_{--}\\
G_{++}-G_{-+}+G_{+-}-G_{--}, & G_{++}-G_{-+}-G_{+-}+G_{--}
\end{pmatrix}.
\end{equation}
In deriving Eq. \eqref{eq:3.23}, we have assumed that the spacetime metric is diagonal.

Combining Eqs. \eqref{eq:3.19} and \eqref{eq:3.24}, it is straightforward to show that
\begin{equation}
\label{eq:3.25}
G_{++}+G_{-+}+G_{+-}+G_{--}=0,
\end{equation}
\begin{equation}
\label{eq:3.26}
\mathbf{G}_{12}^{\circ}(x,y)=2\Theta(y_{0}-x_{0})\big(\mathrm{Tr}\big(\rho_{\chi}\chi(y)\chi(x)\big)-\mathrm{Tr}\big(\rho_{\chi}\chi(x)\chi(y)\big)\big),
\end{equation}
\begin{equation}
\label{eq:3.27}
\mathbf{G}_{21}^{\circ}(x,y)=2\Theta(x_{0}-y_{0})\big(\mathrm{Tr}\big(\rho_{\chi}\chi(x)\chi(y)\big)-\mathrm{Tr}\big(\rho_{\chi}\chi(y)\chi(x)\big)\big)=\mathbf{G}_{12}^{\circ}(y,x),
\end{equation}
\begin{equation}
\label{eq:3.28}
\mathbf{G}_{22}^{\circ}(x,y)=2\mathrm{Tr}\big(\rho_{\chi}[\phi(x),\phi(y)]\big).
\end{equation}
Here, $x_{0}$ and $y_{0}$ denote the temporal components of  $x$ and $y$, respectively, and $\Theta(x)$ is the Heaviside step function. Equation \eqref{eq:3.25} is universally valid for  general open quantum field systems and reflects the conservation of probability in quantum mechanics~\cite{AK,LMS}.

Recalling that the advanced Green function $G_{A}(x,y)$, the retarded Green function $G_{R}(x,y)$, and the Keldysh Green function $G_{K}(x,y)$ are defined as~\cite{JR,AK}
\begin{equation}
\label{eq:3.29}
G_{A}(x,y)=i\Theta(y_{0}-x_{0})\big(\mathrm{Tr}\big(\rho_{\chi}\chi(x)\chi(y)\big)-\mathrm{Tr}\big(\rho_{\chi}\chi(y)\chi(x)\big)\big),
\end{equation}
\begin{equation}
\label{eq:3.30}
G_{R}(x,y)=i\Theta(x_{0}-y_{0})\big(\mathrm{Tr}\big(\rho_{\chi}\chi(y)\chi(x)\big)-\mathrm{Tr}\big(\rho_{\chi}\chi(x)\chi(y)\big)\big),
\end{equation}
 and
\begin{equation}
\label{eq:3.31}
G_{K}(x,y)=-i\mathrm{Tr}\big(\rho_{\chi}[\phi(x),\phi(y)]\big),
\end{equation}
respectively. Combining Eqs. \eqref{eq:3.25}-\eqref{eq:3.31}, one finds that
\begin{equation}
\label{eq:3.32}
\mathbf{G^{\circ}}(x,y)\equiv 2i
\begin{pmatrix}
0, & G_{A}(x,y)\\
G_{R}(x,y), & G_{K}(x,y)
\end{pmatrix}.
\end{equation}
Thus, the matrix Green function $\mathbf{G^{\circ}}(x,y)$ is fully determined by the advanced,  retarded, and  Keldysh Green functions. The relationship between the advanced and  retarded Green functions is $G_{A}(x,y)=G_{R}(y,x)$~\cite{EAI}.  While $G_{A}(x,y)$ and $G_{R}(x,y)$ encode the spectrum information of the field $\chi$,  the Keldysh Green function $G_{K}(x,y)$ encodes its statistical properties~\cite{LMS}.

Additionally, it is useful to note that the Keldysh action  $S_{\phi}^{K}$ can be written as
\begin{eqnarray}\begin{split}
\label{eq:3.33}
&\int d^{4}x \sqrt{-g(x)}\big\{g^{\mu\mu}\partial_{\mu}\phi_{q}\partial_{\mu}\phi_{cl}-m_{\phi}^{2}\phi_{q}\phi_{cl}\big\} \\&=\int d^{4}x \sqrt{-g(x)}\big\{\nabla^{\mu}(\phi_{q}\nabla_{\mu}\phi_{cl})-\phi_{q}\nabla^{\mu}\nabla_{\mu}\phi_{cl}-m_{\phi}^{2}\phi_{q}\phi_{cl}\big\},
\end{split}
\end{eqnarray}
where $\nabla_{\mu}$ denotes the covariant derivative.  On the right hand side of Eq. \eqref{eq:3.33}, the first term can be converted into a boundary term, which does not affect  the dynamics and can thus  be neglected (or equivalently, it can be canceled by introducing a corresponding boundary term). Then, the Keldysh action $S_{\phi}^{K}$ becomes
\begin{eqnarray}\begin{split}
\label{eq:3.34}
S_{\phi}^{K}=-\int d^{4}x \sqrt{-g(x)}\big\{\phi_{q}\nabla^{\mu}\nabla_{\mu}\phi_{cl}+m_{\phi}^{2}\phi_{q}\phi_{cl}\big\}.
\end{split}
\end{eqnarray}
By taking $\delta S_{\phi}^{K}/\delta \phi_{q}=0$, one obtains the dynamical equation
 \begin{equation}
\label{eq:3.35}
\nabla^{\mu}\nabla_{\mu}\phi_{cl}+m_{\phi}^{2}\phi_{cl}=0.
\end{equation}
This is the conventional Klein-Gordon equation, which describes the  dynamics of a scalar field in curved spacetime. Therefore, the Keldysh action in Eq. \eqref{eq:3.34} can indeed be used to describe the dynamics of the scalar field.

Substituting Eqs. \eqref{eq:3.32} and \eqref{eq:3.34} into Eq. \eqref{eq:3.23}, and using the relation $G_{A}(x,y)=G_{R}(y,x)$,  the effective Keldysh action $S_{eff}^{K}$ becomes
\begin{eqnarray}\begin{split}
\label{eq:3.36}
S_{eff}^{K}=&-\int d^{4}x \sqrt{-g(x)}\big\{\phi_{q}\nabla^{\mu}\nabla_{\mu}\phi_{cl}+m_{\phi}^{2}\phi_{q}\phi_{cl}\big\} \\&-\frac{\alpha^{2}}{2}\int d^{4}x d^{4}y \sqrt{-g(x)}\sqrt{-g(y)}\big\{2\phi_{q}(x)\phi_{cl}(y)G_{R}(x,y)+\phi_{q}(x)\phi_{q}(y)G_{K}(x,y)\big\}.
\end{split}
\end{eqnarray}
In the semiclassical region, $\phi_{q}$ is a small quantity, so  higher order terms in $\phi_{q}$ can be neglected~\cite{AK}. In this case, Eq. \eqref{eq:3.36} reduces to
\begin{eqnarray}\begin{split}
\label{eq:3.37}
S_{eff}^{K}=&-\int d^{4}x \sqrt{-g(x)}\big\{\phi_{q}\nabla^{\mu}\nabla_{\mu}\phi_{cl}+m_{\phi}^{2}\phi_{q}\phi_{cl}\big\} \\&-\alpha^{2}\int d^{4}x d^{4}y \sqrt{-g(x)}\sqrt{-g(y)}\phi_{q}(x)\phi_{cl}(y)G_{R}(x,y).
\end{split}
\end{eqnarray}
Equation \eqref{eq:3.37} shows that, in the semiclassical region,  the Keldysh Green function $G_{K}(x,y)$ does not contribute to the effective Keldysh action $S_{eff}^{K}$.

Taking $\delta S_{eff}^{K}/\delta \phi_{q}=0$, one obtains the effective dynamical equation~\cite{AK,LMS} for the variable $\phi_{cl}$, which is given by
\begin{equation}
\label{eq:3.38}
\nabla^{\mu}\nabla_{\mu}\phi_{cl}(x)+m_{\phi}^{2}\phi_{cl}(x)+\alpha^{2}\int d^{4}y \sqrt{-g(y)}\phi_{cl}(y)G_{R}(x,y)=0.
\end{equation}
Equation \eqref{eq:3.38} is an integro-differential equation, which differs from the conventional  Klein-Gordon equation \eqref{eq:3.35}. The additional term in Eq. \eqref{eq:3.38} represents the influence of the bath.  It is evident that when $\alpha\rightarrow 0$, Eq. \eqref{eq:3.38} reduces to Eq. \eqref{eq:3.35}. Similarly, by taking $\delta S_{eff}^{K}/\delta \phi_{cl}=0$, one can obtain the effective dynamical equation for the variable $\phi_{q}$; see Appendix~\ref{sec:B} for details. In the semiclassical region, the dominant contribution  comes from $\phi_{cl}$~\cite{AK}. Thus, in this work, we focus exclusively on the dynamics of $\phi_{cl}$.

In Appendix~\ref{sec:B}, we present an alternative method for deriving the effective dynamical equation \eqref{eq:3.38}.  Although  Appendix~\ref{sec:A} assumes a small coupling constant, the derivation in Appendix~\ref{sec:B} is valid  for both strong and weak interactions. Therefore, Eq. \eqref{eq:3.38} holds for arbitrary coupling strengths. Moreover, the computations in this section do not specify the form of the spacetime metric. Consequently, Eq. \eqref{eq:3.38} is valid not only for a black hole background spacetime but also for  general stationary curved spacetimes.

\section{Reduction of the effective dynamical equation}
\label{sec:4}

The four dimensional integro-differential equation \eqref{eq:3.38} is difficult to solve. To simplify this dynamical equation, we focus on the case in which the coupling constant is small and
the masses of both  the HRP and the field $\chi$ are set to  zero. The spacetime is chosen to be the Schwarzschild spacetime, whose metric is
\begin{equation}
\label{eq:4.1}
ds^{2}=-(1-\frac{2M}{r})dt^{2}+(1-\frac{2M}{r})^{-1}dr^{2}+r^{2}(d\theta^{2}+\mathrm{sin}^{2}\theta d\varphi^{2}).
\end{equation}
Here, $M$ denotes the mass of the black hole.

When $\alpha$ is small,  Eq. \eqref{eq:3.38} can be  solved perturbatively. We expand the field $\phi_{cl}(x)$  as
\begin{equation}
\label{eq:4.2}
\phi_{cl}(x)=\phi_{cl}^{(0)}(x)+\phi_{cl}^{(1)}(x)+\phi_{cl}^{(2)}(x)+\cdot\cdot\cdot,
\end{equation}
where $\phi_{cl}^{(n)}(x)$ ($n=0,1,2,\cdot\cdot\cdot$) represents the  $n$-th order  term. Substituting Eq. \eqref{eq:4.2} into Eq. \eqref{eq:3.38}, we obtain the dynamical equations at each order. The zeroth order equation is
\begin{equation}
\label{eq:4.3}
\nabla^{\mu}\nabla_{\mu}\phi_{cl}^{(0)}=0,
\end{equation}
while the first order  equation takes the form
\begin{equation}
\label{eq:4.4}
\nabla^{\mu}\nabla_{\mu}\phi_{cl}^{(1)}(x)+\alpha^{2}\int d^{4}y \sqrt{-g(y)}\phi_{cl}^{(0)}(y)G_{R}(x,y)=0.
\end{equation}

We approximate the field as $\phi_{cl}(x)\approx\phi_{cl}^{(0)}(x)+\phi_{cl}^{(1)}(x)$.  Adding  Eq. \eqref{eq:4.3} and \eqref{eq:4.4} yields
\begin{equation}
\label{eq:4.5}
\nabla^{\mu}\nabla_{\mu}\phi_{cl}(x)+\alpha^{2}\int d^{4}y \sqrt{-g(y)}\phi_{cl}^{(0)}(y)G_{R}(x,y)=0.
\end{equation}
Note that Eq. \eqref{eq:4.5} becomes an inhomogeneous  differential equation for the field $\phi_{cl}(x)$.
In quantum field theory, the quantity $\phi(x)$  should be interpreted as the scalar field operator. However, in the semiclassical region, when studying black hole radiation via the quantum tunneling approach, $\phi_{cl}(x)$  is often approximately  interpreted as a wavefunction~\cite{EVD,RB,KT,STK,ZJ}. For simplicity, we adopt this interpretation.

In Ref.~\cite{TR}, Damour and Ruffini show that,  as $r\rightarrow 2M$ or $r\rightarrow\infty$, the outgoing wave solution of Eq. \eqref{eq:4.3} is
\begin{eqnarray}
\label{eq:4.6}\phi_{cl}^{(0)}(x)=\frac{1}{\sqrt{4\pi \omega}\:r}
 \begin{cases}
  e^{-i\omega (t-r_{\star}+i4\pi M)}, \quad r<2M,      \\
  e^{-i\omega (t-r_{\star})} , \quad r>2M,
  \end{cases}.
\end{eqnarray}
where $\omega$ is the frequency of the HRP,   and $r_{\star}$ is the tortoise coordinate defined by
\begin{equation}
\label{eq:4.7}
r_{\star}=r+2M\:\mathrm{ln}|\frac{r-2M}{2M}|.
\end{equation}
In Appendix~\ref{sec:C}, we show that when the black hole mass $M$ is small, the outgoing wave function \eqref{eq:4.6} is  approximately valid throughout the entire spacetime region. Thus, in the following, we  set $M$ to be a small quantity.  In the method of Damour and Ruffini, the black hole radiation rate is completely  determined by the outgoing wave component, so the  ingoing wave component does not need to be considered. Equation \eqref{eq:4.6} indicates that when the outgoing wave  passes though the horizon, the Schwarzschild radial coordinate $r$ requires an analytic continuation. The factor $i4\pi M$ in Eq. \eqref{eq:4.6} arises precisely from this analytic continuation~\cite{TR,ZJ}.

For a free scalar field, the retarded Green function at zero temperature coincides with that at finite temperature~\cite{NP}. This conclusion holds in both  flat and curved spacetime. In our model, however,  the fields $\chi$ and $\phi$ interact. Thus, strictly speaking,  $\chi$ cannot be regarded as a free scalar field. Nevertheless, when the coupling constant $\alpha$ is small, the Dyson equation of the Green function~\cite{JR} implies that interaction induced corrections are of higher order. Therefore, for simplicity, these higher order corrections can be neglected. Consequently, in our model, the retarded Green function $G_{R}(x,y)$ may be approximated by that of  a  free scalar field $\chi$ at zero temperature.

The equation of motion for the zero temperature retarded Green function $G_{R}(x,y)$ of a free scalar field  is~\cite{NP}
\begin{equation}
\label{eq:4.10}
\nabla^{\mu}\nabla_{\mu}G_{R}(x,y)=-\frac{1}{\sqrt{-g(x)}}\delta (x-y).
\end{equation}
When the black hole mass $M\rightarrow 0$, the Schwarzschild spacetime reduces to Minkowski spacetime. In this limit,    the retarded Green function solving Eq. \eqref{eq:4.10}  is~\cite{EAI,ME}
\begin{equation}
\label{eq:4.11}
G_{R}(x,y)\big|_{M=0}=\frac{1}{4\pi}\frac{\delta(x_{0}-y_{0}-|\vec{x}-\vec{y}|)}{|\vec{x}-\vec{y}|}.
\end{equation}
Here, $G_{R}(x,y)\big|_{M=0}$  denotes the retarded Green function in the limit $M=0$. The notation  $\vec{x}$ refers to the spatial components of $x$, with $|\vec{x}|\equiv \sqrt{x_{1}^{2}+x_{2}^{2}+x_{3}^{2}}$.   The quantities  $\vec{y}$ and $|\vec{y}|$ are defined analogously.

For a small but nonzero  black hole mass $M$, it is natural to expect that the retarded Green function solving  Eq. \eqref{eq:4.10} can be written as
\begin{equation}
\label{eq:4.11a}
G_{R}(x,y)=G_{R}(x,y)\big|_{M=0}+o_{1}(M).
\end{equation}
The second term on the right hand side of Eq. \eqref{eq:4.11a} represents the correction due to the black hole. Since $M$ is small, this term should be small compared with $G_{R}(x,y)\big|_{M=0}$. In weak gravitational fields, several studies have verified that the retarded Green function indeed takes the form of Eq. \eqref{eq:4.11a}~\cite{ME,YKG}. In the strong field region, to our knowledge, the explicit form of the retarded Green function has not yet been determined. Fortunately, the detailed form of the term $o_{1}(M)$ is not important for our analysis. The reason is that when both  the black hole mass $M$ and the coupling constant $\alpha$ are small, Eqs. \eqref{eq:4.5} and \eqref{eq:4.11a} show that the contribution of $o_{1}(M)$ to the dynamical equation \eqref{eq:4.5} is of higher order  and can therefore be neglected.

For convenience, we define
\begin{equation}
\label{eq:4.13}
F(x)\equiv\alpha^{2}\int d^{4}y \sqrt{-g(y)}\phi_{cl}^{(0)}(y)G_{R}(x,y).
\end{equation}
Substituting Eqs. \eqref{eq:4.6} and \eqref{eq:4.11a} into Eq. \eqref{eq:4.13}, neglecting  higher order terms,  and performing the integration over the variable $y_{0}$,  the function $F(x)$ in Schwarzschild spacetime \eqref{eq:4.1} can be written as
\begin{eqnarray}\begin{split}
\label{eq:4.14}
F(x)=\mathfrak{F}_{1}(x)+\mathfrak{F}_{2}(x),
\end{split}
\end{eqnarray}
where
\begin{eqnarray}\begin{split}
\label{eq:4.15}
\mathfrak{F}_{1}(x)\equiv \frac{\alpha^{2}}{(4\pi)^{\frac{3}{2}}\omega^{\frac{1}{2}}} e^{4\pi M\omega}e^{-i\omega t}\int_{0}^{2M}d|\vec{y}|\int_{0}^{\pi} d \theta'\int_{0}^{2\pi}d\varphi'\mathrm{sin}\theta'\frac{e^{i\omega|\vec{x}-\vec{y}|}}{|\vec{x}-\vec{y}|}|\vec{y}|e^{i\omega |\vec{y}|_{\star}},
\end{split}
\end{eqnarray}
and
\begin{eqnarray}\begin{split}
\label{eq:4.16}
\mathfrak{F}_{2}(x)\equiv  \frac{\alpha^{2}}{(4\pi)^{\frac{3}{2}}\omega^{\frac{1}{2}}}e^{-i\omega t}\int_{2M}^{\infty}d|\vec{y}|\int_{0}^{\pi} d \theta'\int_{0}^{2\pi}d\varphi'\mathrm{sin}\theta'\frac{e^{i\omega|\vec{x}-\vec{y}|}}{|\vec{x}-\vec{y}|}|\vec{y}|e^{i\omega |\vec{y}|_{\star}}.
\end{split}
\end{eqnarray}
In Eqs. \eqref{eq:4.15} and \eqref{eq:4.16}, the tortoise coordinate $|\vec{y}|_{\star}$ is
\begin{equation}
\label{eq:4.17}
|\vec{y}|_{\star}=|\vec{y}|+2M\mathrm{ln}|\frac{|\vec{y}|-2M}{2M}|.
\end{equation}

To reduce the dynamical equation \eqref{eq:4.5}, we first need to simplify the functions $\mathfrak{F}_{1}(x)$ and $\mathfrak{F}_{2}(x)$. Note that the factor $\mathrm{exp}\{i\omega|\vec{x}-\vec{y}|\}/|\vec{x}-\vec{y}|$ in Eqs. \eqref{eq:4.15} and \eqref{eq:4.16} can be expanded as~\cite{GH,TT}
\begin{equation}
\label{eq:4.18}
\frac{e^{i\omega|\vec{x}-\vec{y}|}}{|\vec{x}-\vec{y}|}=4\pi i \omega\sum_{l=0}^{\infty}\sum_{m=-l}^{l}\mathbf{j}_{l}(\omega r_{<})\mathbf{h}_{l}^{(1)}(\omega r_{>})\mathbf{Y}_{lm}^{*}(\theta',\varphi')\mathbf{Y}_{lm}(\theta,\varphi),
\end{equation}
where $r_{<}\equiv min \{|\vec{x}|,|\vec{y}|\}$ and $r_{>}\equiv max \{|\vec{x}|,|\vec{y}|\}$. Here,  $\mathbf{j}_{l}(x)$ is the spherical Bessel function, $\mathbf{h}_{l}^{(1)}(x)$ is spherical Hankel function of the first kind,  $\mathbf{Y}_{lm} (\theta,\varphi)$ is the spherical harmonics, and $\mathbf{Y}_{lm}^{*}$ denotes its complex conjugate.

Substituting Eq. \eqref{eq:4.18} into Eqs. \eqref{eq:4.15} and \eqref{eq:4.16}, the functions $\mathfrak{F}_{1}(x)$ and $\mathfrak{F}_{2}(x)$ become
\begin{eqnarray}\begin{split}
\label{eq:4.19}
\mathfrak{F}_{1}(x)= & i\alpha^{2}\sqrt{\frac{\omega}{4\pi}} e^{4\pi M\omega}e^{-i\omega t}\int_{0}^{2M}d|\vec{y}|\int_{0}^{\pi} d \theta'\int_{0}^{2\pi}d\varphi'\mathrm{sin}\theta'|\vec{y}|e^{i\omega |\vec{y}|_{\star}}\\&\times\sum_{l=0}^{\infty}\sum_{m=-l}^{l}\mathbf{j}_{l}(\omega r_{<})\mathbf{h}_{l}^{(1)}(\omega r_{>})\mathbf{Y}_{lm}^{*}(\theta',\varphi')\mathbf{Y}_{lm}(\theta,\varphi),
\end{split}
\end{eqnarray}
\begin{eqnarray}\begin{split}
\label{eq:4.20}
\mathfrak{F}_{2}(x)= & i\alpha^{2}\sqrt{\frac{\omega}{4\pi}}e^{-i\omega t}\int_{2M}^{\infty}d|\vec{y}|\int_{0}^{\pi} d \theta'\int_{0}^{2\pi}d\varphi'\mathrm{sin}\theta'|\vec{y}|e^{i\omega |\vec{y}|_{\star}}\\&\times\sum_{l=0}^{\infty}\sum_{m=-l}^{l}\mathbf{j}_{l}(\omega r_{<})\mathbf{h}_{l}^{(1)}(\omega r_{>})\mathbf{Y}_{lm}^{*}(\theta',\varphi')\mathbf{Y}_{lm}(\theta,\varphi).
\end{split}
\end{eqnarray}
Using the formula
\begin{equation}
\label{eq:4.21}
\int_{0}^{\pi}d \theta \int_{0}^{2\pi}d \varphi\: \mathrm{sin}\theta \mathbf{Y}_{lm}(\theta,\varphi)=\sqrt{4\pi}\delta_{l0}\delta_{m0},
\end{equation}
the integrations over the variables $\theta'$ and $\varphi'$ in Eqs. \eqref{eq:4.19} and \eqref{eq:4.20} can be performed, and the functions $\mathfrak{F}_{1}(x)$ and $\mathfrak{F}_{2}(x)$ simplify to
\begin{eqnarray}\begin{split}
\label{eq:4.22}
\mathfrak{F}_{1}(x)=  \frac{\alpha^{2}}{\omega\sqrt{4\pi\omega}} e^{4\pi M\omega}e^{-i\omega t}\int_{0}^{2M}d|\vec{y}| \:|\vec{y}|e^{i\omega |\vec{y}|_{\star}}\:\frac{\mathrm{sin}(\omega r_{<})}{r_{<}}\frac{e^{i\omega r_{>}}}{r_{>}},
\end{split}
\end{eqnarray}
\begin{eqnarray}\begin{split}
\label{eq:4.23}
\mathfrak{F}_{2}(x)= \frac{\alpha^{2}}{\omega\sqrt{4\pi\omega}} e^{-i\omega t}\int_{2M}^{\infty}d|\vec{y}| \:|\vec{y}|e^{i\omega |\vec{y}|_{\star}}\:\frac{\mathrm{sin}(\omega r_{<})}{r_{<}}\frac{e^{i\omega r_{>}}}{r_{>}}.
\end{split}
\end{eqnarray}
In deriving Eqs. \eqref{eq:4.22} and \eqref{eq:4.23}, we have used the relations $\mathbf{Y}_{00}=1/\sqrt{4\pi}$, $\mathbf{j}_{0}(x)=\mathrm{sin}(x)/x$,  and $\mathbf{h}_{0}^{(1)}(x)=-ie^{ix}/x$.

Substituting   Eq. \eqref{eq:4.17} into Eqs. \eqref{eq:4.22} and \eqref{eq:4.23}, and performing  the integration over the variable $|\vec{y}|$,  the functions  $\mathfrak{F}_{1}(x)$ and $\mathfrak{F}_{2}(x)$ can be expressed as follows:
\\(\romannumeral1) For $r=|\vec{x}|<2M$:
\begin{eqnarray}\begin{split}
\label{eq:4.24}
\mathfrak{F}_{1}(x)=&C_{1}(x,M)(2M)^{-2i\omega M}\frac{1}{2i}\Big\{e^{4i\omega M}(2i\omega)^{-1-2i\omega M}\big\{\mathbf{\Gamma_{L}}(1+2i\omega M,4i\omega M)\\&\quad-\mathbf{\Gamma_{L}}(1+2i\omega M,2i\omega (2M-r))\big\}-\frac{(2M)^{1+2i\omega M}-(2M-r)^{1+2i\omega M}}{1+2i\omega M}\Big\}\\&+C_{2}(x,M)(2M)^{-2i\omega M}e^{4i\omega M}(2i\omega)^{-1-2i\omega M}\mathbf{\Gamma_{L}}(1+2i\omega M,2i\omega(2M-r)),
\end{split}
\end{eqnarray}
\begin{eqnarray}\begin{split}
\label{eq:4.25}
\mathfrak{F}_{2}(x)=C_{3}(x)(2M)^{-2i\omega M}\mathbf{\Gamma}(1+2i\omega M)(2\omega)^{-1-2i\omega M}\:\mathrm{exp}\{4i\omega M+\frac{i}{2}\pi-\pi\omega M\}.
\end{split}
\end{eqnarray}
\\(\romannumeral2) For $r=|\vec{x}|>2M$:
\begin{eqnarray}\begin{split}
\label{eq:4.26}
\mathfrak{F}_{1}(x)=C_{1}(x,M)(2M)^{-2i\omega M}\frac{1}{2i}\Big\{e^{4i\omega M}\frac{\mathbf{\Gamma_{L}}(1+2i\omega M,4i\omega M)}{(2i\omega)^{1+2i\omega M}}-\frac{(2M)^{1+2i\omega M}}{1+2i\omega M}\Big\},
\end{split}
\end{eqnarray}
\begin{eqnarray}\begin{split}
\label{eq:4.27}
\mathfrak{F}_{2}(x)=&C_{3}(x)(2M)^{-2i\omega M}e^{4i\omega M}(-2i\omega)^{-1-2i\omega M}\mathbf{\Gamma_{U}}(1+2i\omega M,-2i\omega(r-2M))\\&+C_{4}(x)(2M)^{-2i\omega M}\frac{1}{2i}\Big\{e^{4i\omega M}\frac{\mathbf{\Gamma_{L}}(1+2i\omega M,-2i\omega(r-2M))}{(-2i\omega)^{1+2i\omega M}}-\frac{(r-2M)^{1+2i\omega M}}{1+2i\omega M}\Big\}.
\end{split}
\end{eqnarray}
Here, the coefficients $C_{1}(x,M)$, $C_{2}(x,M)$, $C_{3}(x)$, and $C_{4}(x)$ are defined as
\begin{equation}
\label{eq:4.28}
C_{1}(x,M)\equiv\frac{\alpha^{2}}{\omega\sqrt{4\pi\omega}}e^{4\pi M \omega}e^{-i\omega t}\:\frac{e^{i\omega r}}{r},
\end{equation}
\begin{equation}
\label{eq:4.29}
C_{2}(x,M)\equiv\frac{\alpha^{2}}{\omega\sqrt{4\pi\omega}}e^{4\pi M \omega}e^{-i\omega t}\:\frac{\mathrm{sin}(\omega r)}{r},
\end{equation}
\begin{equation}
\label{eq:4.30}
C_{3}(x)\equiv\frac{\alpha^{2}}{\omega\sqrt{4\pi\omega}}e^{-i\omega t}\:\frac{\mathrm{sin}(\omega r)}{r},
\end{equation}
\begin{equation}
\label{eq:4.31}
C_{4}(x)\equiv\frac{\alpha^{2}}{\omega\sqrt{4\pi\omega}}e^{-i\omega t}\:\frac{e^{i\omega r}}{r}.
\end{equation}
In Eqs. \eqref{eq:4.24}-\eqref{eq:4.27}, $\mathbf{\Gamma}(x)$, $\mathbf{\Gamma_{L}}(a,x)$, and $\mathbf{\Gamma_{U}}(a,x)$ represent the Gamma function, the lower incomplete Gamma function and the upper incomplete Gamma function, respectively. They are defined as~\cite{GH}
\begin{equation}
\label{eq:4.32}
\mathbf{\Gamma} (x)\equiv\int_{0}^{\infty}dt e^{-t}t^{x-1},
\end{equation}
\begin{equation}
\label{eq:4.33}
\mathbf{\Gamma_{L}}(a,x)\equiv \int_{0}^{x}dt e^{-t}t^{a-1},
\end{equation}
\begin{equation}
\label{eq:4.34}
\mathbf{\Gamma_{U}}(a,x)\equiv \int_{x}^{\infty}dt e^{-t}t^{a-1}.
\end{equation}
The detailed derivations of Eqs. \eqref{eq:4.24}-\eqref{eq:4.27} are presented in Appendix~\ref{sec:D}.

Equations \eqref{eq:4.24}-\eqref{eq:4.27} show that the form of the function $F(x)$ in curved spacetime is complicated. In the limit  $M\rightarrow 0$, the function $F(x)$ reduces to
\begin{equation}
\label{eq:4.35}
F(x)\big|_{M=0}=\frac{iC_{3}(x)}{2\omega}e^{2i\omega r}+\frac{C_{4}(x)}{4\omega}(1+2i\omega r-e^{2i\omega r}).
\end{equation}
The derivation of Eq. \eqref{eq:4.35} is presented in Appendix~\ref{sec:E}. Based on Eq. \eqref{eq:4.35}, it is straightforward to show that the magnitude of  $F(x)$ in the case  $M=0$ is
\begin{equation}
\label{eq:4.36}
|F(x)|\big|_{M=0}=\frac{\alpha^{2}}{4\sqrt{\pi}\omega^{\frac{3}{2}}}.
\end{equation}
Equation \eqref{eq:4.36} indicates that $|F(x)|\big|_{M=0}$ is independent of the spacetime coordinates.

When studying black hole radiation, the $s$-wave approximation is often adopted~\cite{MF}. In this limit, the wavefunction is spherically symmetric. As a result, the term $\nabla^{\mu}\nabla_{\mu}\phi_{cl}(x)$ in the Schwarzschild spacetime  becomes
\begin{eqnarray}\begin{split}
\label{eq:4.37}
\nabla^{\mu}\nabla_{\mu}\phi_{cl}(x)=&-(1-\frac{2M}{r})^{-1}\partial_{t}^{2}\phi_{cl}(x)+(1-\frac{2M}{r})\partial_{r}^{2}\phi_{cl}(x)\\&
+\frac{2}{r}(1-\frac{2M}{r})\partial_{r}\phi_{cl}(x)+\partial_{r}(1-\frac{2M}{r})\cdot\partial_{r}\phi_{cl}(x).
\end{split}
\end{eqnarray}
In the semiclassical region,  up to the zeroth order in  the Planck constant, one can verify that the term $\nabla^{\mu}\nabla_{\mu}\phi_{cl}(x)$ can be approximated as
\begin{equation}
\label{eq:4.38}
\nabla^{\mu}\nabla_{\mu}\phi_{cl}(x)\approx g^{\mu\nu}\partial_{\mu}\partial_{\nu}\phi_{cl}(x)= -(1-\frac{2M}{r})^{-1}\partial_{t}^{2}\phi_{cl}(x)+(1-\frac{2M}{r})\partial_{r}^{2}\phi_{cl}(x).
\end{equation}
The demonstration of  Eq. \eqref{eq:4.38} is provided in Appendix~\ref{sec:C}.  Under this approximation, the solution to the Klein-Gordon equation \eqref{eq:4.3} is~\cite{KT,RB}
\begin{equation}
\label{eq:4.39}
\phi_{cl}^{0}(x)\propto \mathrm{exp}\big\{-i\omega(t\pm\int_{0}^{r}(1-\frac{2M}{r'})^{-1}dr' )\big\},
\end{equation}
where the symbols ``+" and ``-" correspond to the ingoing and outgoing wavefunctions, respectively. To obtain the correct black hole radiation information, the radial coordinate in Eq. \eqref{eq:4.39}  also requires an analytic continuation when the wavefunction passes though the horizon~\cite{KT,RB,EVD,ETD}.

The outgoing wave solution in Eq. \eqref{eq:4.39} differs from the Damour-Ruffini wavefunction \eqref{eq:4.6}. Thus, if one use the Damour-Ruffini method to study black hole radiation, the approximation in Eq. \eqref{eq:4.38} may not suffice. In Appendix~\ref{sec:C}, we show that in order to be consistent with the wavefunction \eqref{eq:4.6}, the term $\nabla^{\mu}\nabla_{\mu}\phi_{cl}(x)$ should  be approximated as
\begin{equation}
\label{eq:4.40}
\nabla^{\mu}\nabla_{\mu}\phi_{cl}(x)\approx-(1-\frac{2M}{r})^{-1}\partial_{t}^{2}\phi_{cl}(x)+(1-\frac{2M}{r})\partial_{r}^{2}\phi_{cl}(x)+\frac{2}{r}(1-\frac{2M}{r})\partial_{r}\phi_{cl}(x).
\end{equation}
That is, the third term on the right hand side of Eq. \eqref{eq:4.37} should not be neglected. Substituting Eqs. \eqref{eq:4.13} and \eqref{eq:4.40} into Eq. \eqref{eq:4.5}, the dynamical equation \eqref{eq:4.5} simplifies to
\begin{equation}
\label{eq:4.41}
-(1-\frac{2M}{r})^{-1}\partial_{t}^{2}\phi_{cl}(x)+(1-\frac{2M}{r})\partial_{r}^{2}\phi_{cl}(x)+\frac{2}{r}(1-\frac{2M}{r})\partial_{r}\phi_{cl}(x)+F(x)=0.
\end{equation}

Since the Schwarzschild spacetime \eqref{eq:4.1} is stationary, the wavefunction solution of the Klein-Gordon equation \eqref{eq:4.3} can generally be written in the form~\cite{EVD}
\begin{equation}
\label{eq:4.42}
\phi_{cl}^{(0)}(x)\propto e^{-iS(x)}=\mathrm{exp}\big\{-i(\omega t+\mathfrak{S}_{0}(r))\big\}=R^{(0)}(r)e^{-i\omega t}.
\end{equation}
In Eq. \eqref{eq:4.42}, $S(x)$ represents the action of the particle, $\mathfrak{S}_{0}(r)$ is a function of the radial coordinate $r$, and  $R^{(0)}(r)$ denotes the radial wavefunction. One can easily verify that both Eqs. \eqref{eq:4.6} and \eqref{eq:4.39} satisfy the form given in  Eq. \eqref{eq:4.42}.

In this study, although  a bath is present, the spacetime remains stationary. Thus, it is natural to expect that the wavefunction of a particle with  frequency $\omega$ in our model can also be expressed as:
\begin{equation}
\label{eq:4.43}
\phi_{cl}(x)=\frac{1}{\sqrt{4\pi}}R(r)e^{-i\omega t}.
\end{equation}
Here, $R(r)$ represents the radial wavefunction. In Eq. \eqref{eq:4.43}, we have explicitly included the factor $1/\sqrt{4\pi}$, which may be absorbed into the definition of $R(r)$. Note that $1/\sqrt{4\pi}=\mathbf{Y}_{00}(\theta,\varphi)$. Thus, under the $s$-wave approximation, the angular part of the wave function is represented by $1/\sqrt{4\pi}$. Nevertheless, this factor  has no impact on the qualitative results and is therefore  unimportant for the present study.     Substituting Eq. \eqref{eq:4.43} into Eq. \eqref{eq:4.41}, the dynamical equation simplifies to
\begin{eqnarray}\begin{split}
\label{eq:4.45}
\partial_{r}^{2}R(r)+\frac{2}{r}\partial_{r}R(r)+\omega^{2}(1-\frac{2M}{r})^{-2}R(r)+\sqrt{4\pi}(1-\frac{2M}{r})^{-1}e^{i\omega t}\:F(x)=0.
\end{split}
\end{eqnarray}
It is clear that the dynamical equation \eqref{eq:4.45} is a one dimensional inhomogeneous differential equation. It describes the semiclassical dynamics of the HRP. In the case  $\alpha=0$, Eq.\eqref{eq:4.45} corresponds to Eq. \eqref{eq:4.3}.

\section{ Numerical results}
\label {sec:5}

We have reduced the four dimensional  integro-differential equation \eqref{eq:3.38} to a one dimensional inhomogeneous differential equation. However, the function $F(x)$ in Eq. \eqref{eq:4.45} is complicated, making it difficult to  solve the reduced dynamical equation \eqref{eq:4.45} analytically. Therefore, in this section, we numerically study the solution of Eq. \eqref{eq:4.45}.

\begin{figure}[tbp]
\centering
\includegraphics[width=11cm]{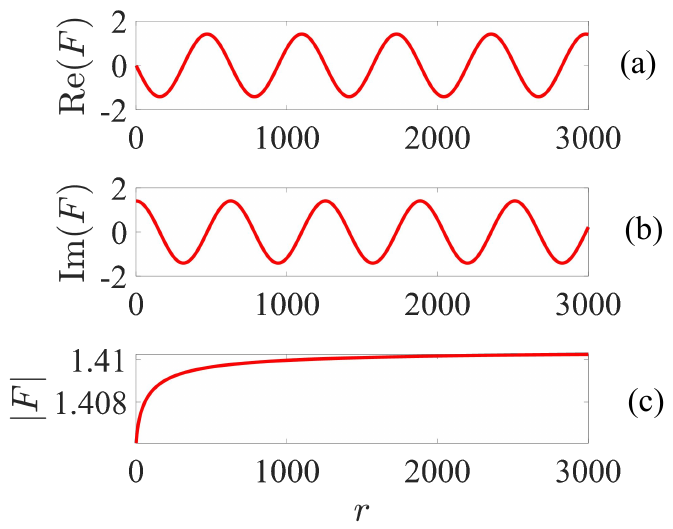}
\caption{\label{fig:1}  Variations of the function $F(x)$ with respect to the Schwarzschild radial coordinate $r$. The horizontal axes of panels (a), (b) and (c) correspond to the  coordinate $r$. The vertical axes, labeled  $Re(F)$, $Im(F)$, and $|F|$,  represent the real part, the imaginary part, and the magnitude of the function $F(x)$, respectively. The parameters are set to: $M=0.1$, $\omega=0.01$, $\alpha=0.1$, and $t=1$. }
\end{figure}

\begin{figure}[tbp]
\centering
\includegraphics[width=10cm]{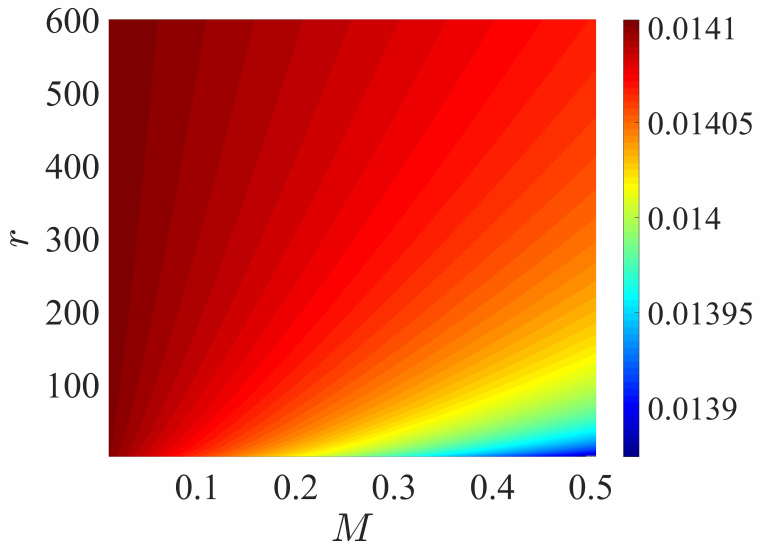}
\caption{\label{fig:2}  Variations of the function $F(x)$. The horizontal and vertical axes represent the black hole mass $M$ and the Schwarzschild radial coordinate $r$, respectively. Different colors represent different values of $|F|$. The parameters are set to: $\omega=0.01$, $\alpha=0.01$, and $t=1$. }
\end{figure}

First, we numerically examine the characteristics of the function $F(x)$, which represents the impact of the bath on the HRP. The analytical form of $F(x)$ is given by Eqs. \eqref{eq:4.14} and \eqref{eq:4.24}-\eqref{eq:4.27}. Figures ~\ref{fig:1} and ~\ref{fig:2} depict the variations of  $F(x)$ with respect to certain variables. Figures ~\ref{fig:1}(a) and ~\ref{fig:1}(b) show that the real  and  imaginary parts of $F(x)$ exhibit characteristics of periodic oscillations, while Figs. ~\ref{fig:1}(c) and  ~\ref{fig:2} indicate that the magnitude of $F(x)$ (denoted as $|F(x)|$)  increases monotonically with the Schwarzschild radial coordinate.  Figure~\ref{fig:2} also shows that  $|F(x)|$ decreases as the black hole mass $M$ increases.

In Fig.~\ref{fig:1}, the black hole mass is taken as $M=0.1$. Thus, the black hole horizon is located at $r=0.2$. In the region $r>200$, we have $\frac{2M}{r}<10^{-3}$, which implies that this region can be approximately regarded as $r=\infty$.   Figure~\ref{fig:1}(c) shows that  as $r\rightarrow \infty$, $|F(x)|$ approaches  a constant. The region $r\rightarrow \infty$ corresponds to  Minkowski spacetime. Equation \eqref{eq:4.36} indicates that in flat spacetime, $|F(x)|$ is  constant. Therefore, the result in Fig.~\ref{fig:1}(c) is consistent with Eq. \eqref{eq:4.36}.

To numerically solve the differential Eq. \eqref{eq:4.45}, one needs to specify the boundary conditions. Recall that in the case without a bath ($\alpha=0$) , the characteristics of  Schwarzschild black hole radiation are determined by the horizon and are independent of the boundary conditions of Eq. \eqref{eq:4.45}. Thus,  when numerically studying  black hole radiation in this case,  the boundary conditions for  Eq. \eqref{eq:4.45} can be set freely. We therefore arbitrarily chose them as $R(r=0)=1+1i$ and $\partial_{r}R(r=0)=-0.1-0.1i$. For the case where $\alpha\neq 0$, to study the effect of the bath on the HRP,   it is convenient  to set the boundary conditions of Eq. \eqref{eq:4.45} to be the same as in the case where $\alpha=0$. When the wavefunction $R(r)$ crosses the horizon, an analytic continuation is also required, which  depends on the characteristics of the horizon.   Since we have neglected the impact of the bath on the spacetime metric, we can adopt the analytic continuation used by Damour and Ruffini, as shown in Eq. \eqref{eq:C13}.

\begin{figure}[tbp]
\centering
\includegraphics[width=12cm]{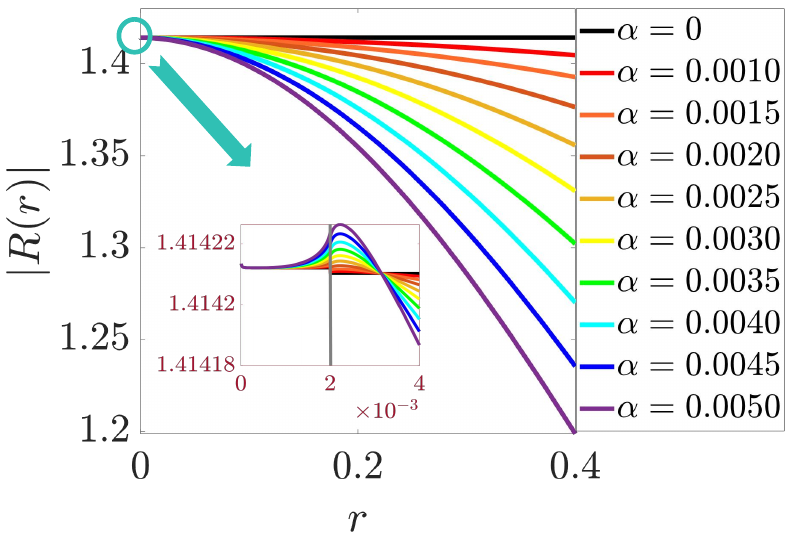}
\caption{\label{fig:3}  Variations of the radial wavefunction over the Schwarzschild radial coordinate $r$. The horizontal and vertical axes represent the coordinate $r$ and the magnitude of radial wavefunction $R(r)$, respectively. Different curves correspond to different values of the coupling constant $\alpha$.  The parameters are set to: $M=0.001$, $\omega=0.0001$, and $t=1$. }
\end{figure}

\begin{figure}[tbp]
\centering
\includegraphics[width=8cm]{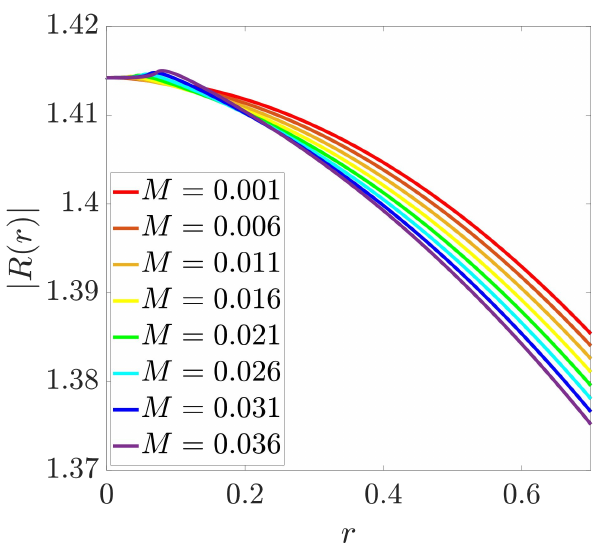}
\caption{\label{fig:4}   Variations of the radial wavefunction with respect to the Schwarzschild radial coordinate $r$. The horizontal and vertical axes represent the coordinate $r$ and the magnitude of radial wavefunction $R(r)$, respectively. Different curves correspond to different values of the black hole mass $M$.  The parameters are set to: $\alpha=0.001$, $\omega=0.0001$, and $t=1$. }
\end{figure}

Figures~\ref{fig:3} and ~\ref{fig:4} depict the numerical solutions of the dynamical equation \eqref{eq:4.45}.  The horizontal axes of Figs.~\ref{fig:3} and ~\ref{fig:4} represent the Schwarzschild radial coordinate $r$, while the vertical axes represent the  magnitude of the radial wavefunction, denoted as $|R(r)|$. To ensure the accuracy of the numerical results, the parameter values used in Figs.~\ref{fig:3} and ~\ref{fig:4} are smaller than those in Figs.~\ref{fig:1} and ~\ref{fig:2}. Under these parameters, it is easy to verify numerically  that the oscillation period of  the real (or imaginary) part of the function $F(x)$ is on the order of $10^{5}$, while the size of the black hole horizon is on the order of $10^{-3}$ to $10^{-2}$. Thus, in this case, the oscillation period of  $F(x)$ can be considered effectively  infinite, and its oscillatory behavior   can be neglected.

In Fig.~\ref{fig:3}, the curves in the small box show a magnified view of the region near the black hole horizon taken from the curves in the large box.  The vertical gray line in the small box marks the location of the horizon.  Different curves correspond to different coupling constants. The black curve corresponds to the case with $\alpha=0$, while the other curves correspond to  $\alpha\neq 0$. In other words, the black curve represents the radial wavefunction when no bath is present, whereas the colored curves include the effect of the bath. Hawking's prediction for black hole radiation  corresponds to the black curve.   The black hole mass in Fig.~\ref{fig:3} is taken to be $M=0.001$, and  the horizon is located at $r=0.002$. At $r=0.4$, we have $\frac{2M}{r}\ll 1$, indicating  that an observer located at $r=0.4$ can be approximately regarded as being  far from the black hole. In addition, smaller values of $|R(r)|$ correspond to a smaller probability of finding a particle at  position $r$; thus, smaller $|R(r)|$ indicates weaker black hole radiation.    Therefore, Fig.~\ref{fig:3} shows that when a black hole is surrounded by a thermal bath, a distant observer detects less black hole radiation than in the case without a bath. As the coupling constant increases, the black hole radiation is more strongly suppressed. Furthermore, it is well known that in the absence of a bath,  a larger black hole mass corresponds to  weaker black hole radiation. Figure~\ref{fig:4} shows that even when the black hole is surrounded by a bath, this results remains valid.

Note that the effective dynamical equation of a particle in the Caldeira-Leggett model is the Langevin equation~\cite{AA}. The model given in Eq. \eqref{eq:2.1} is a curved spacetime version of the Caldeira-Leggett model~\cite{SSS}. This implies that the non-relativistic limit of the model in Eq. \eqref{eq:2.1} should be consistent with the Caldeira-Leggett model. Consequently, the non-relativistic limit of the effective dynamical equation in Eq. \eqref{eq:3.38} should be consistent with the Langevin equation. In particular, the non-relativistic limit of the third term on the left hand side of Eq. \eqref{eq:3.38} should correspond to the friction force in the Langevin equation. In the Caldeira–Leggett model with the bath at zero temperature, the friction force suppresses particle tunneling~\cite{AA,JR}. In the weak coupling limit, we have shown that the temperature effects of the bath can be neglected in the model given in Eq. \eqref{eq:2.1}.  Therefore, analogous to the Caldeira–Leggett model, it is reasonable to infer that the effect of the third term on the left hand side of Eq. \eqref{eq:3.38} may suppress the tunneling of a particle out of a black hole. This inference is consistent with the  numerical results presented in Fig.~\ref{fig:3}.

\section{Conclusion and discussion}
\label{sec:6}

In this work, we investigated  the impact of a thermal bath surrounding a black hole on its Hawking radiation. Neglecting  internal degrees of freedom, both the bath and the HRP can be described by a scalar field.    The coupling between the HRP and the thermal bath is assumed to be bilinear.    The spacetime background is taken to be the Schwarzschild spacetime. In this stationary spacetime, one can define the Hamiltonian.  For simplicity, we neglect the impact of the thermal bath on the spacetime metric.

Following the framework  of open quantum systems, we treat the HRP as the system and the thermal bath as the environment. The dynamical information of the system is encoded in its reduced density matrix. To derive the effective dynamical equation, we express the reduced density matrix in the path integral form. After tracing out the  bath degrees of freedom, we obtain the effective Keldysh action for the system, which can be used to describe the nonequilibrium evolution of the system. In this effective Keldysh action, the influence of the bath is encoded in the matrix Green function, which is determined by the advanced, retarded, and Keldysh Green functions. Taking the variation of the effective Keldysh action yields the effective dynamical equation. In the semiclassical region, the Keldysh Green function does not contribute to the this equation.

We have shown that the dynamical equation \eqref{eq:3.38} is valid for both  strong and weak coupling. Equation \eqref{eq:3.38} is a four dimensional integro-differential equation. When the coupling constant $\alpha\rightarrow 0$, it reduces to the conventional Klein-Gordon equation.
Equation \eqref{eq:3.38} is difficult to solve. To simplify this equation,  we focus on the case of small coupling.  In this limit,  the four dimensional integro-differential equation reduces to a four dimensional  inhomogeneous differential equation, where the inhomogeneous term represents the effect of the bath on the HRP dynamics.     For convenience, we introduce a function $F(x)$ to denotes this term. To further simplify the analysis, we assume that  the black hole mass is small and that both scalar fields are massless. Under these assumptions, the retarded Green function can be approximated by a simple form, as shown in Eq. \eqref{eq:4.11a}. Consequently, the dynamical equation reduces to a one dimensional inhomogeneous differential equation.

We find that  $|F(x)|$ increases monotonically with the Schwarzschild radial coordinate, whereas in Minkowski spacetime, $|F(x)|$ is  constant. We also find that $|F(x)|$  decreases as the black hole mass increases. Moreover,  we show that both  $Re(F(x))$ and $Im(F(x))$ exhibit periodic oscillatory behavior. For the numerical solution of the reduced dynamical equation, we demonstrated that the boundary conditions can be chosen arbitrarily for the purpose of studying Hawking radiation. Our results show that an observer located
far  from the black hole will  detect weaker radiation when the black hole is surrounded by a thermal bath, and the radiation becomes further suppressed as the coupling constant increases. We also show that the black hole radiation decreases as the black hole mass increases, and this conclusion holds regardless of whether a bath is present.

\section*{Acknowledgements}
Hong Wang  was supported by the National Natural Science Foundation of China Grant No.12234019. Hong Wang thanks for the help from Professor Erkang Wang.

\appendix
\section{Derivation of Eq. \eqref{eq:3.18} }
\label{sec:A}

According to Eqs. \eqref{eq:3.10} and \eqref{eq:3.13}, the time discretized version of the influence functional is
\begin{equation}
\label{eq:A1}
\mathrm{Tr}_{\chi}\Big\{\mathrm{exp}\big\{-i\delta t \sum_{n}(H_{\chi}+H_{int}(\phi_{+(n-1)}))\big\}\rho_{\chi}(-\infty)\mathrm{exp}\big\{i\delta t \sum_{n}(H_{\chi}+H_{int}(\phi_{-(n+1)}))\big\}\Big\}.
\end{equation}
Using the identity $\mathrm{Tr}(ABC)=\mathrm{Tr}(CAB)$ and the first-order Trotter-Suzuki decomposition~\cite{SEC}
\begin{equation}
\label{eq:A2}
\mathrm{exp}\big\{-i\delta t (H_{\chi}+H_{int})\big\}=\mathrm{exp}\big\{-i\delta t H_{\chi}\big\}\mathrm{exp}\big\{-i\delta t H_{int}\big\}+o(\delta t^{2}),
\end{equation}
Eq. \eqref{eq:A1} can be simplified to
\begin{equation}
\label{eq:A3}
\mathrm{Tr}_{\chi}\Big\{\mathrm{exp}\big\{i\delta t \sum_{n}(H_{int}(\phi_{-(n+1)})-H_{int}(\phi_{+(n-1)}))\big\}\rho_{\chi}(-\infty)\Big\}.
\end{equation}

Returning to the continuous limit, the influence functional in Eq. \eqref{eq:A3} becomes
\begin{equation}
\label{eq:A4}
I(\phi_{+},\phi_{-})=\mathrm{Tr}_{\chi}\Big\{\mathrm{exp}\big\{i\int dt \sum_{n}(H_{int}(\phi_{-})-H_{int}(\phi_{+}))\big\}\rho_{\chi}(-\infty)\Big\}.
\end{equation}
When the coupling constant is small, Eq. \eqref{eq:A4} admits  perturbative expansion~\cite{JR}:
\begin{eqnarray}\begin{split}
\label{eq:A5}
I(\phi_{+},\phi_{-})=\mathrm{Tr}_{\chi}\Big\{&\rho_{\chi}\big\{1+i\int dt H_{int}(\phi_{-})-\frac{1}{2}\int dt_{1} dt_{2}\mathbf{\tilde{T}}\big(H_{int}(\phi_{-},t_{2})H_{int}(\phi_{-},t_{1})\big)+\cdot\cdot\cdot\big\}\\&\times\big\{1-i\int dt H_{int}(\phi_{+})-\frac{1}{2}\int dt_{1} dt_{2}\mathbf{T}\big(H_{int}(\phi_{+},t_{1})H_{int}(\phi_{+},t_{2})\big)+\cdot\cdot\cdot\big\}\Big\}.
\end{split}
\end{eqnarray}
For bilinear coupling, when the bath is in thermal equilibrium,  the expectation value of the interaction Hamiltonian operator vanishes~\cite{MBD}. For other types of couplings,  this is also a standard  assumption  in  open quantum system theory~\cite{HF,HJ2}.  Using this fact, and substituting Eq. \eqref{eq:2.7} into Eq. \eqref{eq:A5}, the influence functional becomes
\begin{eqnarray}\begin{split}
\label{eq:A6}
I(\phi_{+},\phi_{-})=&1+\alpha^{2}\mathrm{Tr}_{\chi}\Big\{\rho_{\chi}\int d^{4}x_{1}\sqrt{-g(x_{1})}\phi_{-}(x_{1})\chi(x_{1})\int d^{4}x_{2}\sqrt{-g(x_{2})}\phi_{+}(x_{2})\chi(x_{2})\Big\}\\&-\frac{\alpha^{2}}{2}\mathrm{Tr}_{\chi}\Big\{\rho_{\chi}\int d^{4}x_{1}d^{4}x_{2}\mathbf{\tilde{T}}\Big\{\sqrt{-g(x_{2})}\phi_{-}(x_{2})\chi(x_{2})\sqrt{-g(x_{1})}\phi_{-}(x_{1})\chi(x_{1})\Big\}\Big\}\\&-\frac{\alpha^{2}}{2}\mathrm{Tr}_{\chi}\Big\{\rho_{\chi}\int d^{4}x_{1}d^{4}x_{2}\mathbf{T}\Big\{\sqrt{-g(x_{1})}\phi_{+}(x_{1})\chi(x_{1})\sqrt{-g(x_{2})}\phi_{+}(x_{2})\chi(x_{2})\Big\}\Big\}\\&+\cdot\cdot\cdot.
\end{split}
\end{eqnarray}

Using the formula
\begin{equation}
\label{eq:A7}
\int_{b}^{a}f(x)dx\int_{b}^{a}h(x)dx=\frac{1}{2}\int_{b}^{a}\int_{b}^{a}dxdy\big\{f(x)h(y)+f(y)h(x)\big\},
\end{equation}
Eq. \eqref{eq:A6} can be rewritten as
\begin{eqnarray}\begin{split}
\label{eq:A8}
I(\phi_{+},\phi_{-})=&1+\frac{\alpha^{2}}{2}\mathrm{Tr}_{\chi}\Big\{\rho_{\chi}\int d^{4}x_{1}d^{4}x_{2}\sqrt{-g(x_{1})}\sqrt{-g(x_{2})}\phi_{-}(x_{1}) \phi_{+}(x_{2})\chi(x_{1})\chi(x_{2})\Big\}\\&+\frac{\alpha^{2}}{2}\mathrm{Tr}_{\chi}\Big\{\rho_{\chi}\int d^{4}x_{1}d^{4}x_{2}\sqrt{-g(x_{1})}\sqrt{-g(x_{2})}\phi_{-}(x_{2})\phi_{+}(x_{1})\chi(x_{2}) \chi(x_{1})\Big\}\\&-\frac{\alpha^{2}}{2}\mathrm{Tr}_{\chi}\Big\{\rho_{\chi}\int d^{4}x_{1}d^{4}x_{2}\sqrt{-g(x_{1})}\sqrt{-g(x_{2})}\phi_{-}(x_{1})\phi_{-}(x_{2})\mathbf{\tilde{T}}\big(\chi(x_{2})\chi(x_{1})\big)\Big\}\\&-\frac{\alpha^{2}}{2}\mathrm{Tr}_{\chi}\Big\{\rho_{\chi}\int d^{4}x_{1}d^{4}x_{2}\sqrt{-g(x_{1})}\sqrt{-g(x_{2})}\phi_{+}(x_{1})\phi_{+}(x_{2})\mathbf{T}\big(\chi(x_{1})\chi(x_{2})\big)\Big\}\\&+\cdot\cdot\cdot.
\end{split}
\end{eqnarray}
Substituting the definitions  in Eq. \eqref{eq:3.19} into Eq. \eqref{eq:A8}, the influence functional  becomes
\begin{eqnarray}\begin{split}
\label{eq:A9}
I(\phi_{+},\phi_{-})=1-\frac{\alpha^{2}}{2}\int d^{4}x_{1}d^{4}x_{2}\sqrt{-g(x_{1})}\sqrt{-g(x_{2})}\vec{\phi}(x_{1})\mathbf{G}(x_{1},x_{2})\vec{\phi}^{\mathrm{T}}(x_{2})+\cdot\cdot\cdot,
\end{split}
\end{eqnarray}
which has the same structure  as  Eq. (A6) in Ref.~\cite{MBD}.  The authors  of~\cite{MBD} showed  that  influence functionals of this form satisfy
\begin{eqnarray}\begin{split}
\label{eq:A10}
I(\phi_{+},\phi_{-})=\mathrm{exp}\Big\{-\frac{\alpha^{2}}{2}\int d^{4}x_{1}d^{4}x_{2}\sqrt{-g(x_{1})}\sqrt{-g(x_{2})}\vec{\phi}(x_{1})\mathbf{G}(x_{1},x_{2})\vec{\phi}^{\mathrm{T}}(x_{2})\Big\}.
\end{split}
\end{eqnarray}
Replacing $x_{1}$ and $x_{2}$ with $x$ and $y$, respectively, Eq. \eqref{eq:A10} becomes Eq. \eqref{eq:3.18}. We emphasize  that the effective Keldysh action \eqref{eq:3.36} and the effective dynamical equation \eqref{eq:3.38} are both derived from  the influence functional in Eq. \eqref{eq:A10}. They may also  be obtained  by alternative  methods, as shown in Appendix~\ref{sec:B}, which further  confirms the correctness  of Eq. \eqref{eq:A10}.

\section{Alternative derivation of the effective dynamical equation \eqref{eq:3.38} }
\label{sec:B}

The Keldysh action of the total system is
\begin{eqnarray}\begin{split}
\label{eq:B1}
S^{K}_{tot}=&\frac{1}{2}\int d^{4}x_{1} \sqrt{-g} \big\{g^{\mu\mu}(\partial_{\mu}\phi_{+})^{2}-g^{\mu\mu}(\partial_{\mu}\phi_{-})^{2}-m_{\phi}^{2}(\phi_{+}^{2}-\phi_{-}^{2})\big\}\\&+\frac{1}{2}\int d^{4}x_{1} \sqrt{-g} \big\{g^{\mu\mu}(\partial_{\mu}\chi_{+})^{2}-g^{\mu\mu}(\partial_{\mu}\chi_{-})^{2}-m_{\chi}^{2}(\chi_{+}^{2}-\chi_{-}^{2})\big\}\\&+\alpha\int d^{4}x_{1} \sqrt{-g}(\phi_{+}\chi_{+}-\phi_{-}\chi_{-}).
\end{split}
\end{eqnarray}
On the right hand side of Eq. \eqref{eq:B1}, the first line is the Keldysh action of the free scalar field $\phi$ in curved spacetime, which is equivalent to Eq. \eqref{eq:3.15}. The second line gives the Keldysh action of the field $\chi$, while the third line represents the Keldysh action of the interaction term.

The Keldysh action of the scalar field $\chi$ can be written as
\begin{eqnarray}\begin{split}
\label{eq:B2}
&\frac{1}{2}\int d^{4}x_{1} \sqrt{-g} \big\{g^{\mu\mu}(\partial_{\mu}\chi_{+})^{2}-g^{\mu\mu}(\partial_{\mu}\chi_{-})^{2}-m_{\chi}^{2}(\chi_{+}^{2}-\chi_{-}^{2})\big\}\\&=-\frac{1}{2}\int d^{4}x_{1} \sqrt{-g} \big\{\chi_{+}(\nabla^{\mu}\nabla_{\mu}+m_{\chi}^{2})\chi_{+}-\chi_{-}(\nabla^{\mu}\nabla_{\mu}+m_{\chi}^{2})\chi_{-}\big\}.
\end{split}
\end{eqnarray}
Defining~\cite{NP}
\begin{equation}
\label{eq:B3}
\mathbb{D}(x_{1},x_{2})\equiv (\nabla^{\mu}\nabla_{\mu}+m_{\chi}^{2})\frac{\delta(x_{1}-x_{2})}{\sqrt{-g(x_{2})}},
\end{equation}
the Keldysh action of $\chi$ can be expressed  as
\begin{eqnarray}\begin{split}
\label{eq:B4}
&-\frac{1}{2}\int d^{4}x_{1} \sqrt{-g(x_{1})} \big\{\chi_{+}(\nabla^{\mu}\nabla_{\mu}+m_{\chi}^{2})\chi_{+}-\chi_{-}(\nabla^{\mu}\nabla_{\mu}+m_{\chi}^{2})\chi_{-}\big\}\\&=-\frac{1}{2}\int d^{4}x_{1} d^{4}x_{2}\sqrt{-g(x_{1})} \sqrt{-g(x_{2})} \chi_{+}(x_{1})\mathbb{D}(x_{1},x_{2})\chi_{+}(x_{2})\\&\quad+\frac{1}{2}\int d^{4}x_{1} d^{4}x_{2}\sqrt{-g(x_{1})} \sqrt{-g(x_{2})} \chi_{-}(x_{1})\mathbb{D}^(x_{1},x_{2})\chi_{-}(x_{2}).
\end{split}
\end{eqnarray}

The Keldysh rotation for the scalar field $\phi$ is defined in Eq. \eqref{eq:3.22}. Similarly,  for the field $\chi$,
\begin{eqnarray}
\label{eq:B5}
 \begin{cases}
       \chi_{cl}\equiv\frac{1}{\sqrt{2}}(\chi_{+}+\chi_{-}) \\
       \chi_{q}\equiv\frac{1}{\sqrt{2}}(\chi_{+}-\chi_{-})
  \end{cases}.
\end{eqnarray}
After performing the Keldysh rotation on $\chi$, and following  Eq. \eqref{eq:B4}, we assume that the Keldysh action of the free scalar field $\chi$ in curved spacetime can be written as
\begin{eqnarray}
\label{eq:B6}
 \frac{1}{2}\int d^{4}x_{1} d^{4}x_{2}\sqrt{-g(x_{1})} \sqrt{-g(x_{2})} (\chi_{cl}(x_{1}),\chi_{q}(x_{1}))\mathbf{K}^{-1}(x_{1},x_{2}) (\chi_{cl}(x_{2}),\chi_{q}(x_{2}))^{\mathrm{T}}.
\end{eqnarray}
Here, $\mathbf{K}(x_{1},x_{2})$ is defined as
\begin{equation}
\label{eq:B7}
\mathbf{K}(x_{1},x_{2})\equiv
\begin{pmatrix}
K^{K}(x_{1},x_{2}), & K^{R}(x_{1},x_{2})\\
K^{A}(x_{1},x_{2}), & 0
\end{pmatrix},
\end{equation}
where $K^{K}(x_{1},x_{2})$, $K^{R}(x_{1},x_{2})$, and $K^{A}(x_{1},x_{2})$ are unspecified  functions to be determined later.   The matrix Green function of a nonequilibrium  quantum system typically  takes this form~\cite{JR,AK,LMS}.

Performing the Keldysh rotation on both  $\phi$ and $\chi$, Eq. \eqref{eq:B1} becomes
\begin{eqnarray}\begin{split}
\label{eq:B8}
S^{K}_{tot}=&-\int d^{4}x_{1} \sqrt{-g(x_{1})}\big\{\phi_{q}(x_{1})(\nabla^{\mu}\nabla_{\mu}+m_{\phi}^{2})\phi_{cl}(x_{1})\big\}\\& +\frac{1}{2}\int d^{4}x_{1} d^{4}x_{2}\sqrt{-g(x_{1})} \sqrt{-g(x_{2})} (\chi_{cl}(x_{1}),\chi_{q}(x_{1}))\mathbf{K}^{-1}(x_{1},x_{2}) (\chi_{cl}(x_{2}),\chi_{q}(x_{2}))^{\mathrm{T}}\\&+\alpha \int d^{4}x_{1} \sqrt{-g(x_{1})}\big(\phi_{cl}(x_{1}),\phi_{q}(x_{1})\big)\mathbf{\sigma_{1}} \big(\chi_{cl}(x_{1}),\chi_{q}(x_{1})\big)^{\mathrm{T}},
\end{split}
\end{eqnarray}
where
\begin{equation}
\label{eq:B9}
\mathbf{\sigma_{1}}\equiv
\begin{pmatrix}
0, & 1\\
1, & 0
\end{pmatrix}
\end{equation}
is the first Pauli matrix.

The partition function of the total system is
\begin{equation}
\label{eq:B10}
Z=\int D \phi_{cl}D\phi_{q}D\chi_{cl}D\chi_{q}\:\mathrm{exp}\big\{iS^{K}_{tot}\big\}.
\end{equation}
Substituting Eq. \eqref{eq:B8} into Eq. \eqref{eq:B10} and using the integral formula
\begin{equation}
\label{eq:B11}
\int d^{n}x\:\mathrm{exp}\{-\frac{i}{2}\vec{x}R\vec{x}^{\mathrm{T}}+i\vec{J}\vec{x}^{\mathrm{T}}\}=\sqrt{\frac{(2\pi i)^{n}}{\mathrm{det}R}}\mathrm{exp}\big\{\frac{i}{2}\vec{J}R^{-1}\vec{J}^{T}\big\}
\end{equation}
the integrals over  $\chi_{cl}$ and $\chi_{q}$ yield
\begin{eqnarray}\begin{split}
\label{eq:B12}
Z=\int D \phi_{cl}D\phi_{q}&\:\mathrm{exp}\Big\{-i\int d^{4}x_{1}\sqrt{-g(x_{1})}\phi_{q}(x_{1})(\nabla^{\mu}\nabla_{\mu}+m_{\phi}^{2})\phi_{cl}(x_{1})-\frac{i}{2}\alpha^{2}\int d^{4}x_{1} d^{4}x_{2}\\&\times\sqrt{-g(x_{1})} \sqrt{-g(x_{2})}\big (\phi_{cl}(x_{1}),\phi_{q}(x_{1})\big)\mathbf{\sigma_{1}}\mathbf{K}(x_{1},x_{2})\mathbf{\sigma_{1}}\big (\phi_{cl}(x_{2}),\phi_{q}(x_{2})\big)^{\mathrm{T}}\Big\}.
\end{split}
\end{eqnarray}
In Eq. \eqref{eq:B12}, we have omitted  factors that can be absorbed into the normalization constant.

From Eq. \eqref{eq:B12},  the effective Keldysh action is
\begin{eqnarray}\begin{split}
\label{eq:B13}
S_{eff}^{K}=&-\int d^{4}x_{1}\sqrt{-g(x_{1})}\phi_{q}(x_{1})(\nabla^{\mu}\nabla_{\mu}+m_{\phi}^{2})\phi_{cl}(x_{1})\\&-\frac{1}{2}\alpha^{2}\int d^{4}x_{1} d^{4}x_{2}\sqrt{-g(x_{1})} \sqrt{-g(x_{2})}\big (\phi_{cl}(x_{1}),\phi_{q}(x_{1})\big)\mathbf{\sigma_{1}}\mathbf{K}(x_{1},x_{2})\mathbf{\sigma_{1}}\big (\phi_{cl}(x_{2}),\phi_{q}(x_{2})\big)^{\mathrm{T}}.
\end{split}
\end{eqnarray}
Substituting Eqs. \eqref{eq:B7} and \eqref{eq:B9} into Eq. \eqref{eq:B13}, we obtain
\begin{eqnarray}\begin{split}
\label{eq:B14}
S_{eff}^{K}=&-\int d^{4}x_{1}\sqrt{-g(x_{1})}\phi_{q}(x_{1})(\nabla^{\mu}\nabla_{\mu}+m_{\phi}^{2})\phi_{cl}(x_{1})\\&-\frac{1}{2}\alpha^{2}\int d^{4}x_{1} d^{4}x_{2}\sqrt{-g(x_{1})} \sqrt{-g(x_{2})}\Big\{\phi_{q}(x_{1})K^{R}(x_{1},x_{2})\phi_{cl}(x_{2})\\&\quad\quad+\phi_{q}(x_{1})K^{K}(x_{1},x_{2})\phi_{q}(x_{2})+\phi_{cl}(x_{1})K^{A}(x_{1},x_{2})\phi_{q}(x_{2})\Big\}.
\end{split}
\end{eqnarray}

We emphasize that in deriving Eq. \eqref{eq:B14}, we did not assume  the coupling constant $\alpha$ to be  small. Therefore Eq. \eqref{eq:B14} is valid for both strong and weak coupling. When $\alpha$ is  small, Eq. \eqref{eq:B14} must reduce to Eq. \eqref{eq:3.36}. Imposing this requirement fixes  $K^{K}(x_{1},x_{2})=G_{K}(x_{1},x_{2})$, $K^{R}(x_{1},x_{2})=G_{R}(x_{1},x_{2})$ and $K^{A}(x_{1},x_{2})=G_{A}(x_{1},x_{2})$.
This demonstrates that the form of the effective Keldysh action in Eq. \eqref{eq:3.36} holds  for both strong and weak coupling. Consequently, the resulting effective dynamical equation \eqref{eq:3.38} is also valid  for arbitrary coupling strength.

To derive the effective dynamical equation for the variable $\phi_{q}$, we note that the Keldysh action $S_{\phi}^{K}$ can also be written as
\begin{eqnarray}\begin{split}
\label{eq:B15}
S_{\phi}^{K}&=\int d^{4}x \sqrt{-g(x)}\big\{g^{\mu\mu}\partial_{\mu}\phi_{q}\partial_{\mu}\phi_{cl}-m_{\phi}^{2}\phi_{q}\phi_{cl}\big\} \\&=-\int d^{4}x \sqrt{-g(x)}\big\{\phi_{cl}\nabla^{\mu}\nabla_{\mu}\phi_{q}+m_{\phi}^{2}\phi_{q}\phi_{cl}\big\}.
\end{split}
\end{eqnarray}
Thus, the effective Keldysh action $S_{eff}^{K}$ can be written as
\begin{eqnarray}\begin{split}
\label{eq:B16}
S_{eff}^{K}=&-\int d^{4}x \sqrt{-g(x)}\big\{\phi_{cl}\nabla^{\mu}\nabla_{\mu}\phi_{q}+m_{\phi}^{2}\phi_{q}\phi_{cl}\big\} \\&-\frac{\alpha^{2}}{2}\int d^{4}x d^{4}y \sqrt{-g(x)}\sqrt{-g(y)}\big\{2\phi_{q}(x)\phi_{cl}(y)G_{R}(x,y)+\phi_{q}(x)\phi_{q}(y)G_{K}(x,y)\big\}.
\end{split}
\end{eqnarray}
Taking $\delta S_{eff}^{K}/\delta \phi_{cl}=0$, one  obtains the effective dynamical equation for the variable $\phi_{q}$,
\begin{equation}
\label{eq:B17}
\nabla^{\mu}\nabla_{\mu}\phi_{q}(x)+m_{\phi}^{2}\phi_{q}(x)+\alpha^{2}\int d^{4}y \sqrt{-g(y)}\phi_{q}(y)G_{R}(x,y)=0.
\end{equation}
By solving Eq.\eqref{eq:B17}, one can obtain the dynamical information for $\phi_{q}$.

\section{Approximate outgoing wave solution of Eq. \eqref{eq:4.3} }
\label{sec:C}
For clarity, in this section we explicitly include the Planck constant $\hbar$. The wavefunction solution to the Klein-Gordon equation \eqref{eq:4.3} can be assumed to take the form~\cite{EVD}
\begin{equation}
\label{eq:C1}
\phi_{cl}^{(0)}(x)=e^{-\frac{i}{\hbar}S(x)}.
\end{equation}
According to Eqs. \eqref{eq:4.37} and \eqref{eq:C1}, the term $\nabla^{\mu}\nabla_{\mu}\phi_{cl}^{(0)}(x)$ can be written as
\begin{eqnarray}\begin{split}
\label{eq:C2}
\nabla^{\mu}\nabla_{\mu}\phi_{cl}^{(0)}(x)=&-(1-\frac{2M}{r})^{-1}\Big\{-\frac{i}{\hbar}\partial_{t}^{2}S\: e^{-\frac{i}{\hbar}S}+(\frac{i}{\hbar})^{2}(\partial_{t}S)^{2}e^{-\frac{i}{\hbar}S}\Big\}\\&-\frac{i}{\hbar}\frac{2}{r}(1-\frac{2M}{r})\partial_{r}S\:e^{-\frac{i}{\hbar}S}-
\frac{i}{\hbar}\partial_{r}(1-\frac{2M}{r})\partial_{r}Se^{-\frac{i}{\hbar}S}\\&+(1-\frac{2M}{r})\Big\{-\frac{i}{\hbar}\partial_{r}^{2}S\: e^{-\frac{i}{\hbar}S}+(\frac{i}{\hbar})^{2}(\partial_{r}S)^{2}e^{-\frac{i}{\hbar}S}\Big\}.
\end{split}
\end{eqnarray}
Consequently, the Klein-Gordon equation \eqref{eq:4.3} simplifies to
\begin{eqnarray}\begin{split}
\label{eq:C3}
&i\hbar(1-\frac{2M}{r})^{-1}\partial_{t}^{2}S+(1-\frac{2M}{r})^{-1}(\partial_{t}S)^{2}-i\hbar(1-\frac{2M}{r})\partial_{r}^{2}S\\&
-\Big\{i\hbar\frac{2}{r}(1-\frac{2M}{r})+i\hbar\partial_{r}(1-\frac{2M}{r})\Big\}\partial_{r}S-(1-\frac{2M}{r})(\partial_{r}S)^{2}=0.
\end{split}
\end{eqnarray}

In the semiclassical region, the action $S(x)$ can be expanded as
\begin{equation}
\label{eq:C4}
S(x)=S_{0}(x)+\hbar S_{1}(x)+\hbar^{2}S_{2}(x)+\cdot\cdot\cdot.
\end{equation}
Keeping only the zeroth  order in  $\hbar$, Eq. \eqref{eq:C3} reduces to
\begin{equation}
\label{eq:C5}
(\partial_{t}S_{0})^{2}-(1-\frac{2M}{r})^{2}(\partial_{r}S_{0})^{2}=0.
\end{equation}
Keeping terms to   the first order in  $\hbar$, Eq. \eqref{eq:C3} becomes
\begin{eqnarray}\begin{split}
\label{eq:C6}
&i\hbar(1-\frac{2M}{r})^{-1}\partial_{t}^{2}S_{0}-\Big\{i\hbar\frac{2}{r}(1-\frac{2M}{r})+i\hbar\partial_{r}(1-\frac{2M}{r})\Big\}\partial_{r}S_{0}-i\hbar(1-\frac{2M}{r})\partial_{r}^{2}S_{0}\\&
\quad\quad+(1-\frac{2M}{r})^{-1}(\partial_{t}S_{0})^{2}
-(1-\frac{2M}{r})\Big\{(\partial_{r}S_{0})^{2}+2\hbar\partial_{r}S_{0}\partial_{r}S_{1}\Big\}=0.
\end{split}
\end{eqnarray}

Solving  Eq. \eqref{eq:C5} yields  the wavefunction   in Eq. \eqref{eq:4.39}.  For the outgoing mode,  Eq. \eqref{eq:4.39} gives
\begin{equation}
\label{eq:C7}
S_{0}(x)=\omega t -\omega\int_{0}^{r}(1-\frac{2M}{r'})^{-1}dr'.
\end{equation}
Equation \eqref{eq:C7} implies $\partial_{t}^{2}S_{0}=0$.  For a small  black hole mass $M$, one has $S_{0}(x)\approx\omega t-\omega r+o_{2}(M)$ and $\partial_{r}(1-2M/r)=o_{2}(M)$. Here, $o_{2}(M)$  denotes terms proportional to $M$. Substituting these results into Eq. \eqref{eq:C6} and neglecting  higher order contributions, Eq. \eqref{eq:C6} simplifies  to
\begin{eqnarray}\begin{split}
\label{eq:C8}
-2i\hbar\frac{1}{r}-2\hbar\partial_{r}S_{1}(x)=0.
\end{split}
\end{eqnarray}
It is straightforward to check  that the solution of Eq. \eqref{eq:C8} is
\begin{eqnarray}\begin{split}
\label{eq:C9}
S_{1}(x)=-i\mathrm{ln }(r).
\end{split}
\end{eqnarray}

Substituting Eqs. \eqref{eq:C4}, \eqref{eq:C7}, and \eqref{eq:C9} into Eq. \eqref{eq:C1}, one can show  that for small   $M$  the outgoing wavefunction solution of the Klein-Gordon equation \eqref{eq:4.3} is approximately
\begin{eqnarray}\begin{split}
\label{eq:C10}
\phi_{cl}^{(0)}(x)\approx\mathrm{ exp}\big\{-\frac{i}{\hbar}\big(S_{0}(x)+\hbar S_{1}(x)\big)\big\}=\frac{1}{r}\mathrm{ exp}\big\{-\frac{i}{\hbar}\omega\big( t-\int_{0}^{r}(1-\frac{2M}{r'})^{-1}dr'\big)\big\}.
\end{split}
\end{eqnarray}
Equation \eqref{eq:C10} shows that the factor $1/r$ originates from  the first order contribution  in $\hbar$.
For the Schwarzschild black hole, the  tortoise coordinate $r_{\star}$ can be expressed as~\cite{KT}
\begin{eqnarray}\begin{split}
\label{eq:C11}
r_{\star}=\int_{0}^{r}(1-\frac{2M}{r'})^{-1}dr'.
\end{split}
\end{eqnarray}
Hence the wavefunction \eqref{eq:C10} can be rewritten as
\begin{eqnarray}\begin{split}
\label{eq:C12}
\phi_{cl}^{(0)}(x)=\frac{1}{r}\mathrm{ exp}\big\{-\frac{i}{\hbar}\omega\big( t-r_{\star}\big)\big\}.
\end{split}
\end{eqnarray}

Following the suggestion of Damour and Ruffini, when the wave function crosses the horizon from the inside to the outside, one should apply the following analytic continuation to the radial coordinate $r$~\cite{TR}:
\begin{equation}
\label{eq:C13}
2M-r\rightarrow (r-2M)e^{i\pi}.
\end{equation}
This implies that the wavefunction \eqref{eq:C12} must be modified to~\cite{ZJ}
\begin{eqnarray}
\label{eq:C14}\phi_{cl}^{(0)}(x)=\frac{1}{r}
 \begin{cases}
  \mathrm{exp}\{-\frac{i}{\hbar}\omega (t-r_{\star}+i4\pi M)\}; \quad r<2M      \\
  \mathrm{exp}\{-\frac{i}{\hbar}\omega (t-r_{\star})\} ; \quad r>2M
  \end{cases}.
\end{eqnarray}
Setting  $\hbar=1$, and multiplying by  a factor $1/\sqrt{4\pi\omega}$, Eq. \eqref{eq:C14} becomes Eq. \eqref{eq:4.6}.
The factor $1/\sqrt{4\pi\omega}$ can be absorbed into the normalization constant and does  not  affect the qualitative results, so it is  irrelevant   for our purposes.
In this sense, Eq. \eqref{eq:C14} is equivalent to Eq. \eqref{eq:4.6}.   Since no restriction on $r$ was imposed in deriving Eq. \eqref{eq:C14}, it remains approximately valid throughout the spacetime when $M$ is small.

According to Eqs. \eqref{eq:C2} and \eqref{eq:C5}, the term appearing  in Eq. \eqref{eq:4.37}
\begin{eqnarray}\begin{split}
\label{eq:C15}
-(1-\frac{2M}{r})^{-1}\partial_{t}^{2}\phi_{cl}(x)+(1-\frac{2M}{r})\partial_{r}^{2}\phi_{cl}(x)
\end{split}
\end{eqnarray}
contributes to the function $S_{0}(x)$. Thus,  to  zeroth order in $\hbar$, the term $\nabla^{\mu}\nabla_{\mu}\phi_{cl}(x)$ may be approximated by the expression  in Eq. \eqref{eq:4.38}. Equations \eqref{eq:C2} and \eqref{eq:C8} show  that
\begin{eqnarray}\begin{split}
\label{eq:C16}
(1-\frac{2M}{r})\partial_{r}^{2}\phi_{cl}(x)+\frac{2}{r}(1-\frac{2M}{r})\partial_{r}\phi_{cl}(x)
\end{split}
\end{eqnarray}
contributes to  $S_{1}(x)$. Therefore,  to ensure   consistency with Eq. \eqref{eq:C14}, the term $\nabla^{\mu}\nabla_{\mu}\phi_{cl}(x)$ should be approximated by the expression in Eq. \eqref{eq:4.40}.

\section{Derivation of Eqs. \eqref{eq:4.24}-\eqref{eq:4.27} }
\label{sec:D}

Substituting the definitions of $r_{<}$,  $r_{>}$ and $|\vec{y}|_{\star}$ into Eqs. \eqref{eq:4.22} and \eqref{eq:4.23}, the functions $\mathfrak{F}_{1}(x)$ and  $\mathfrak{F}_{2}(x)$  can be written as follows:
\\ (\romannumeral1) When $r<2M$,
\begin{eqnarray}\begin{split}
\label{eq:D1}
\mathfrak{F}_{1}(x)&=C_{1}(x,M)(2M)^{-2i\omega M}\int_{0}^{r}d|\vec{y}|\:\mathrm{sin}(\omega |\vec{y}|)\:(2M-|\vec{y}|)^{2i\omega M}e^{i\omega |\vec{y}|}\\&\quad\quad+C_{2}(x,M)(2M)^{-2i\omega M}\int_{r}^{2M}d|\vec{y}|\:(2M-|\vec{y}|)^{2i\omega M}e^{2i\omega |\vec{y}|}\\&\equiv\mathfrak{F}_{1}(x)\big|_{r<2M},
\end{split}
\end{eqnarray}
\begin{eqnarray}\begin{split}
\label{eq:D2}
\mathfrak{F}_{2}(x)=C_{3}(x)(2M)^{-2i\omega M}\int_{2M}^{\infty}d|\vec{y}|\:(|\vec{y}|-2M)^{2i\omega M}e^{2i\omega |\vec{y}|}\equiv\mathfrak{F}_{2}(x)\big|_{r<2M}.
\end{split}
\end{eqnarray}
\\ (\romannumeral2) When $r>2M$,
\begin{eqnarray}\begin{split}
\label{eq:D3}
\mathfrak{F}_{1}(x)=&C_{1}(x,M)(2M)^{-2i\omega M}\int_{0}^{2M}d|\vec{y}|\:\mathrm{sin}(\omega |\vec{y}|)\:(2M-|\vec{y}|)^{2i\omega M}e^{i\omega |\vec{y}|}\equiv\mathfrak{F}_{1}(x)\big|_{r>2M},
\end{split}
\end{eqnarray}
\begin{eqnarray}\begin{split}
\label{eq:D4}
\mathfrak{F}_{2}(x)&=C_{3}(x)(2M)^{-2i\omega M}\int_{r}^{\infty}d|\vec{y}|\:(|\vec{y}|-2M)^{2i\omega M}e^{2i\omega |\vec{y}|}\\&\quad\quad+C_{4}(x)(2M)^{-2i\omega M}\int_{2M}^{r}d|\vec{y}|\:\mathrm{sin}(\omega r)\:(|\vec{y}|-2M)^{2i\omega M}e^{i\omega |\vec{y}|}\\&\equiv\mathfrak{F}_{2}(x)\big|_{r>2M}.
\end{split}
\end{eqnarray}
In Eqs. \eqref{eq:D1}-\eqref{eq:D4}, for convenience, we have introduced the symbols $\mathfrak{F}_{1}(x)\big|_{r<2M}$, $\mathfrak{F}_{2}(x)\big|_{r<2M}$, $\mathfrak{F}_{1}(x)\big|_{r>2M}$ and $\mathfrak{F}_{2}(x)\big|_{r>2M}$ to denotes  the functions $\mathfrak{F}_{1}(x)$ and  $\mathfrak{F}_{2}(x)$ in different regions.

We  first  compute $\mathfrak{F}_{1}(x)\big|_{r>2M}$. It is straightforward  to show
\begin{eqnarray}\begin{split}
\label{eq:D5}
&\int_{0}^{2M}d|\vec{y}|\:\mathrm{sin}(\omega |\vec{y}|)\:(2M-|\vec{y}|)^{2i\omega M}e^{i\omega |\vec{y}|}\\&=\frac{1}{2i}\int_{0}^{2M}d|\vec{y}|\:(2M-|\vec{y}|)^{2i\omega M}e^{2i\omega |\vec{y}|}-\frac{1}{2i}\int_{0}^{2M}d|\vec{y}|\:(2M-|\vec{y}|)^{2i\omega M}\\&=\frac{1}{2i}e^{4i\omega M}\int_{0}^{2M}du \:u^{2i\omega M}e^{-2i\omega u}-\frac{(2M)^{1+2i\omega M}}{2i(1+2i\omega M)}.
\end{split}
\end{eqnarray}
According to the definition of the lower incomplete Gamma function in Eq. \eqref{eq:4.33}, one can show that
\begin{equation}
\label{eq:D6}
\int_{0}^{2M}du \:u^{2i\omega M}e^{-2i\omega u}=(2i\omega)^{-1-2i\omega M}\mathbf{\Gamma_{L}}(1+2i\omega M, 4i\omega M).
\end{equation}
Substituting Eqs. \eqref{eq:D5} and \eqref{eq:D6} into Eq. \eqref{eq:D3},  the function $\mathfrak{F}_{1}(x)\big|_{r>2M}$  can be written as
\begin{equation}
\label{eq:D7}
\mathfrak{F}_{1}(x)\big|_{r>2M}=C_{1}(x,M)(2M)^{-2i\omega M}\frac{1}{2i}\Big\{e^{4i\omega M}\frac{\mathbf{\Gamma_{L}}(1+2i\omega M,4i\omega M)}{(2i\omega)^{1+2i\omega M}}-\frac{(2M)^{1+2i\omega M}}{1+2i\omega M}\Big\}.
\end{equation}

Using the same method, one can verify that
\begin{eqnarray}\begin{split}
\label{eq:D8}
&\int_{0}^{r}d|\vec{y}|\:\mathrm{sin}(\omega |\vec{y}|)\:(2M-|\vec{y}|)^{2i\omega M}e^{i\omega |\vec{y}|}\\&=\frac{1}{2i}e^{4i\omega M}(2i\omega )^{-1-2i\omega M}\Big\{\mathbf{\Gamma_{L}}(1+2i\omega M,4i\omega M)-\mathbf{\Gamma_{L}}(1+2i\omega M,2i\omega (2M-r))\Big\}\\&\quad\quad-\frac{1}{2i}(1+2i\omega M)^{-1}\Big\{(2M)^{1+2i\omega M}-(2M-r)^{1+2i\omega M}\Big\},
\end{split}
\end{eqnarray}
and
\begin{eqnarray}\begin{split}
\label{eq:D9}
\int_{r}^{2M}d|\vec{y}|\:(2M-|\vec{y}|)^{2i\omega M}e^{2i\omega |\vec{y}|}&=e^{4i\omega M}\int_{0}^{2M-r}du\: u^{2i\omega M}e^{-2i\omega u}\\&=e^{4i\omega M}\frac{\mathbf{\Gamma_{L}}(1+2i\omega M,2i\omega (2M-r))}{(2i\omega)^{1+2i\omega M}}.
\end{split}
\end{eqnarray}
Substituting Eqs. \eqref{eq:D8} and \eqref{eq:D9} into Eq. \eqref{eq:D1}, one finds that the function $\mathfrak{F}_{1}(x)\big|_{r<2M}$  takes the  form  presented in Eq. \eqref{eq:4.24}.

In Eq. \eqref{eq:D2}, the integral with respect to  $|\vec{y}|$ can be rewritten as
\begin{eqnarray}\begin{split}
\label{eq:D10}
\int_{2M}^{\infty}d|\vec{y}|\:(|\vec{y}|-2M)^{2i\omega M}e^{2i\omega |\vec{y}|}=e^{4i\omega M}\int_{0}^{\infty}du\: u^{2i\omega M}e^{2i\omega u}.
\end{split}
\end{eqnarray}
According to the definition of the Gamma function in Eq. \eqref{eq:4.32}, this integral can be expressed as
\begin{eqnarray}\begin{split}
\label{eq:D11}
\int_{2M}^{\infty}d|\vec{y}|\:(|\vec{y}|-2M)^{2i\omega M}e^{2i\omega |\vec{y}|}=\mathbf{\Gamma} (1+2i\omega M)\frac{\mathrm{exp}\big\{4i\omega M+\frac{i}{2}\pi-\pi\omega M\big\}}{(2\omega)^{1+2i\omega M}}.
\end{split}
\end{eqnarray}
Substituting Eq. \eqref{eq:D11} into Eq. \eqref{eq:D2}, one  obtains  Eq. \eqref{eq:4.25}.

In Eq. \eqref{eq:D4},  the integral
\begin{equation}
\label{eq:D12}
\int_{r}^{\infty}d|\vec{y}|\:(|\vec{y}|-2M)^{2i\omega M}e^{2i\omega |\vec{y}|}
\end{equation}
can be rewritten as
\begin{equation}
\label{eq:D13}
e^{4i\omega M}\int_{r-2M}^{\infty}du\: u^{2i\omega M}e^{2i\omega u}.
\end{equation}
Using the definition of the upper incomplete Gamma function in Eq. \eqref{eq:4.34}, the integral in Eq. \eqref{eq:D13} is equal to
\begin{equation}
\label{eq:D14}
e^{4i\omega M}(-2i\omega)^{-1-2i\omega M}\mathbf{\Gamma_{U}}(1+2i\omega M,-2i\omega(r-2M)).
\end{equation}
Similarly to the case of Eq. \eqref{eq:D5},  one can verify that
\begin{eqnarray}\begin{split}
\label{eq:D15}
&\int_{2M}^{r}d|\vec{y}|\:\mathrm{sin}(\omega |\vec{y}|)\:(|\vec{y}|-2M)^{2i\omega M}e^{i\omega |\vec{y}|}\\&=e^{4i\omega M}\frac{\mathbf{\Gamma_{L}}(1+2i\omega M,-2i\omega(r-2M))}{2i(-2i\omega)^{1+2i\omega M}}-\frac{(r-2M)^{1+2i\omega M}}{2i(1+2i\omega M)}.
\end{split}
\end{eqnarray}
Substituting Eqs. \eqref{eq:D14} and \eqref{eq:D15} into Eq. \eqref{eq:D4},  the function $\mathfrak{F}_{2}(x)\big|_{r>2M}$ can be simplified to the form given in Eq. \eqref{eq:4.27}.

\section{Derivation of Eq. \eqref{eq:4.35} }
\label{sec:E}

According to Eqs. \eqref{eq:4.13}-\eqref{eq:4.16}, in the case of $M=0$, the function $F(x)$ simplifies to
\begin{equation}
\label{eq:E1}
F(x)\big|_{M=0}=\frac{\alpha^{2}}{(4\pi)^{\frac{3}{2}}\omega^{\frac{1}{2}}}e^{-i\omega t}\int_{0}^{\infty} d|\vec{y}|\int_{0}^{2\pi}d\varphi'\int_{0}^{\pi}d\theta'\:\mathrm{sin}\theta'\frac{e^{i\omega|\vec{x}-\vec{y}|}}{|\vec{x}-\vec{y}|}|\vec{y}|e^{i\omega |\vec{y}|}.
\end{equation}
Substituting Eq. \eqref{eq:4.18} into Eq. \eqref{eq:E1},   the function $F(x)\big|_{M=0}$ can be written as
\begin{eqnarray}\begin{split}
\label{eq:E2}
F(x)\big|_{M=0}=&i\alpha^{2}\sqrt{\frac{\omega}{4\pi}}e^{-i\omega t}\int_{0}^{\infty} d|\vec{y}|\int_{0}^{2\pi}d\varphi'\int_{0}^{\pi}d\theta'\:\mathrm{sin}\theta'\:|\vec{y}|e^{i\omega |\vec{y}|}\\&\times\sum_{l=0}^{\infty}\sum_{m=-l}^{l}\mathbf{j}_{l}(\omega r_{<})\mathbf{h}_{l}^{(1)}(\omega r_{>})\mathbf{Y}_{lm}^{*}(\theta',\varphi')\mathbf{Y}_{lm}(\theta,\varphi).
\end{split}
\end{eqnarray}
Using the results   $\mathbf{Y}_{00}=1/\sqrt{4\pi}$, $\mathbf{j}_{0}(x)=\mathrm{sin}(x)/x$,  $\mathbf{h}_{0}^{(1)}(x)=-ie^{ix}/x$ and Eq. \eqref{eq:4.21},  the integration over  $\theta'$ and $\varphi'$ can be performed, yielding
\begin{eqnarray}\begin{split}
\label{eq:E3}
F(x)\big|_{M=0}=\frac{\alpha^{2}}{\omega\sqrt{4\pi\omega}}e^{-i\omega t}\int_{0}^{\infty}d|\vec{y}| \:|\vec{y}|e^{i\omega |\vec{y}|}\:\frac{\mathrm{sin}(\omega r_{<})}{r_{<}}\frac{e^{i\omega r_{>}}}{r_{>}}.
\end{split}
\end{eqnarray}

Using the definitions of $r_{<}$ and $r_{>}$, Eq. \eqref{eq:E3} can be written as
\begin{eqnarray}\begin{split}
\label{eq:E4}
F(x)\big|_{M=0}=&\frac{\alpha^{2}}{\omega\sqrt{4\pi\omega}}e^{-i\omega t}\Big\{\frac{e^{i\omega r}}{r}\int_{0}^{r}d|\vec{y}|\: \mathrm{sin}(\omega |\vec{y}|)e^{i\omega |\vec{y}|}+\frac{\mathrm{sin}(\omega r)}{r}\int_{r}^{\infty}d|\vec{y}|e^{2i\omega |\vec{y}|}\Big\}.
\end{split}
\end{eqnarray}
Using Eqs. \eqref{eq:4.30} and \eqref{eq:4.31}, Eq. \eqref{eq:E4} can be expressed  as
\begin{eqnarray}\begin{split}
\label{eq:E5}
F(x)\big|_{M=0}=&C_{4}(x)\int_{0}^{r}d|\vec{y}|\: \mathrm{sin}(\omega |\vec{y}|)e^{i\omega |\vec{y}|}+C_{3}(x)\int_{r}^{\infty}d|\vec{y}|e^{2i\omega |\vec{y}|}.
\end{split}
\end{eqnarray}
Performing  the integrals  over  $|\vec{y}|$ in Eq. \eqref{eq:E5}, one obtains Eq. \eqref{eq:4.35}.

\end{document}